\documentclass[a4paper]{article}

\usepackage[margin=1in]{geometry}  
\usepackage[english]{babel}
\usepackage[utf8x]{inputenc}
\usepackage[T1]{fontenc}

\usepackage[dvipsnames]{xcolor}
\usepackage{graphicx}
\usepackage{subcaption}
\usepackage{longtable} 
\usepackage{multirow}
\usepackage{listings}
\usepackage{makecell}
\usepackage{array}
\usepackage{float}
\usepackage{dsfont}
\usepackage{rotating}
\usepackage{booktabs}
\usepackage{enumitem}
\usepackage{tikz}
\usepackage{pgf}
\usepackage{xcolor}

\usepackage{amsmath}
\usepackage{amssymb}
\usepackage{amsthm}
\usepackage{bm}
\usepackage{algorithm,algpseudocode}

\usepackage{mathtools}
\usepackage{mathrsfs}

\usepackage{tcolorbox}
\usepackage{wrapfig}

\usepackage{parskip}
\allowdisplaybreaks

\usepackage{natbib}
\usepackage[
  colorlinks,
  citecolor=blue,
  linkcolor=red,
  anchorcolor=red,
  urlcolor=blue
]{hyperref}
\mathtoolsset{showonlyrefs}
\usepackage{authblk}
\usepackage{todonotes}
\theoremstyle{plain}

\newtheorem{theorem}{Theorem}
\newtheorem{lemma}{Lemma}

\newtheorem{remark}{Remark}
\newtheorem{assumption}{Assumption}
\newtheorem{proposition}{Proposition}

\newcommand{\RR}{\mathbb{R}}
\newcommand{\EE}{\mathbb{E}}
\newcommand{\PP}{\mathbb{P}}
\newcommand{\ind}{\mathds{1}}

\newcommand{\fdr}{\textnormal{FDR}}
\newcommand{\fdp}{\textnormal{FDP}}

\newcommand{\argmax}[1]{\underset{#1}{\arg\!\max}}
\newcommand{\argmin}[1]{\underset{#1}{\arg\!\min}}

\usepackage{xspace}

\newcommand{\stepb}[1]{\overset{\rm (b)}{#1}}

\newcommand{\given}{{\,|\,}}
\newcommand{\biggiven}{\,\big{|}\,}

\newcommand{\bigggiven}{\,\bigg{|}\,}

\def\@#1\@{\begin{align}#1\end{align}}
\def\$#1\${\begin{align*}#1\end{align*}}

\definecolor{myblue}{rgb}{.8, .8, 1}
\definecolor{mathblue}{rgb}{0.2472, 0.24, 0.6} 
\definecolor{mathred}{rgb}{0.6, 0.24, 0.442893}
\definecolor{mathyellow}{rgb}{0.6, 0.547014, 0.24}

\newcommand{\tZ}{{\tilde{Z}}}

\newcommand{\cA}{{\mathcal{A}}}

\newcommand{\cC}{{\mathcal{C}}}
\newcommand{\cD}{{\mathcal{D}}}

\newcommand{\cF}{{\mathcal{F}}}
\newcommand{\cG}{{\mathcal{G}}}
\newcommand{\cH}{{\mathcal{H}}}

\newcommand{\cL}{{\mathcal{L}}}

\newcommand{\cN}{{\mathcal{N}}}

\newcommand{\cS}{{\mathcal{S}}}

\newcommand{\cX}{{\mathcal{X}}}
\newcommand{\cY}{{\mathcal{Y}}}

\newcommand{\algoname}{\textnormal{ACS}}
\renewcommand{\hat}{\widehat}
\newcommand{\tr}{\textnormal{train}}
\newcommand{\te}{\textnormal{test}}

\newcommand{\ca}{\textnormal{cal}}
\newcommand{\lab}{\textnormal{lab}}
\newcommand{\one}{\mathbf{1}}
\makeatletter
\newcommand{\printfnsymbol}[1]{%
  \textsuperscript{\@fnsymbol{#1}}%
}
\makeatother
\long\def\comment#1{}

\definecolor{yg}{RGB}{32, 186, 181}

\title{ACS: An interactive framework for conformal selection}
\author[1]{Yu Gui\thanks{Author names listed alphabetically.}}
\affil[1]{Department of Statistics and Data Science, University of Pennsylvania}
\author[1]{Ying Jin\printfnsymbol{1}}
\author[2]{Yash Nair\printfnsymbol{1}}
\affil[2]{Department of Statistics, Stanford University }
\author[1]{Zhimei Ren\printfnsymbol{1}}
\date{\today}

\begin{document}
\maketitle
\begin{abstract}
This paper presents adaptive conformal selection (ACS), an interactive framework for 
model-free selection with guaranteed error control. 
Building on conformal selection~\citep{jin2023selection}, ACS 
generalizes the approach to support human-in-the-loop adaptive data analysis. Under the ACS framework, 
we can partially reuse the data to boost the selection power, make decisions on the fly while 
exploring the data, and incorporate new information or preferences as they arise.
The key to ACS is a carefully designed principle that controls the information 
available for decision making, allowing the data analyst to explore the data adaptively 
while maintaining rigorous control of the false discovery rate (FDR). 
Based on the ACS framework, we provide concrete selection algorithms for various goals, including 
model update/selection, diversified selection, and incorporating newly available labeled data.  
The effectiveness of ACS is demonstrated through extensive numerical simulations and 
real-data applications in large language model (LLM) deployment and drug discovery.
\end{abstract}

\section{Introduction}
\label{sec:intro}
We study the problem of identifying promising units from a potentially 
large pool of candidates---for example, selecting drug candidates with high 
binding affinity to a target from a space of chemical compounds~\citep{huang2007drug,lavecchia2013virtual}, choosing job candidates from a pool of applicants~\citep{sajjadiani2019using}, or filtering trustworthy outputs from language models~\citep{gui2024conformal}. 
This problem can be viewed as selecting units whose unknown labels (e.g., binding affinities) 
obey a desired property.  To this end, powerful machine learning models have been developed to 
predict these unknown label based on observable features (e.g., chemical structure), 
enabling efficient shortlisting in place of costly manual evaluation.
However, due to the inherent uncertainty in predictions from complex models, 
a central challenge 
is to control the errors in the selection process,   
ensuring that the selected candidates are indeed promising.

To address this challenge,~\citet{jin2023selection} proposed conformal selection (CS), a statistically 
rigorous framework for shortlisting promising candidates with the assistance of any prediction machine.
Assuming access to a set of labeled data whose properties of interest are known (e.g., drug candidates 
with measured binding affinities), CS sets aside a portion of them (training data) to fit a prediction model; 
it then uses the remaining labeled data (calibration data) to determine which unlabeled data (test data) to select. 
Provided that the calibration and test data are exchangeable, 
CS guarantees control of the false discovery rate (FDR)---the expected fraction of the selected units whose unknown outcomes do not satisfy the property of interest. 
Such guarantees hold regardless of the prediction model and do not require any modeling assumptions of the data distribution,
thereby providing robust and model-agnostic guarantees for the selection.

Despite its flexibility and robustness, a key limitation of CS is that it separates the model choice from the selection process. 
All the components---including the split of labeled data, the model fitting procedure, 
choice of the conformity score that determines the ordering in which the units are selected---must be fixed a priori. 
This requirement prevents any changes to the analysis plan after seeing the calibration and test data, which can often be too rigid in practice. 
To illustrate this point, we present  
two applications below where the need for adaptive analysis strategies naturally arises:

\vspace{0.5em}
\begin{itemize}
\item[(1)] 
In virtual screening for drug discovery, CS is used to identify which drug candidates should 
advance to later stages of development.
With a pre-specified  conformity score (fitted on the training fold), it ranks the test (unlabeled) drug candidates by their scores and leverages the calibration (labeled) data to determine how many top-ranked test drugs to select. 
However, if adaptive analysis were permitted, a scientist could adjust the initial data split---for example, reallocating some calibration data to the training set if the model appears underfit due to insufficient training samples, or switching to a more data-efficient model class if the original choice performs poorly. This flexibility is especially valuable when labeled data from prior campaigns are scarce. 
Furthermore, instead of adhering to a fixed ranking determined by the pre-specified conformity score,
the scientist may re-prioritize selection after evaluating some candidates,  
in order to improve the overall diversity or novelty of the selection set~\citep{nair2025diversifying}.

\item[(2)] In the context of large language model (LLM) reliability,
CS has been instantiated to filter trustworthy outcomes from a foundation model~\citep{gui2024conformal}, selecting qualified outputs for direct 
deployment in downstream tasks while routing others to human experts 
for review. 
This expert feedback may create new labeled data beyond what was initially available. 
A natural question arises: 
can we leverage the newly labeled test data---when it becomes available---to improve the model fitting without compromising 
the error control guarantee?

\end{itemize}
\vspace{0.5em}

In both applications, the practitioner may want to update the analysis plan after inspecting the data for various purposes. 
However, it is unclear how to achieve this without 
violating the FDR control offered by CS. 
Motivated by this practical need, we aim to address the following question:
\vspace{0.25em}
\begin{quote}
{\it To what extent can we 
incorporate additional information and/or make data-driven decisions within CS in an adaptive fashion, while maintaining the error control?}
\end{quote}
\vspace{0.25em}
Answering this question is nontrivial, since arbitrary changes to the analysis plan may introduce the risk of double-dipping. 
In order to preserve error control, we design a framework that governs precisely what information the decision maker is allowed to use in their analysis at each timestep of the selection process.


\subsection{Adaptive Conformal Selection}
\label{sec:intro_acs}
We propose {\em adaptive conformal selection (ACS)}, a new framework that 
extends CS to support interactive data analysis in the 
selection process while maintaining exact FDR control in finite samples. 
In a nutshell, ACS is a sequential procedure that iteratively refines the ordering of candidates 
until a stopping criterion is met, at which point all the top-ordered candidates are selected. 
This interactive approach enables flexible changes to the analysis plan---allowing
practitioners to incorporate new data, update prediction models, and prioritize complicated metrics like diversity after partially inspecting the data. 

ACS begins by obtaining an initial predictive model with a subset of labeled data,
which is then used to produce an initial ordering---from least to most promising---of both the (remaining) labeled and unlabeled candidates. 
The procedure then screens the least promising unit, reveals its label information, and examines the stopping criterion (i.e., whether an estimated false discovery proportion falls below the target level). 
If the criterion is not yet met, the newly screened data point can be used 
in any way to update the order of the remaining candidates. 
This process continues iteratively, with the candidate ranking updated at each step, until the stopping condition is satisfied. At that point, the remaining unlabeled candidates are selected as the final set.


As we shall show in the paper, ACS enjoys the same FDR control as CS.  To achieve this, the only requirement in ACS is that the rule determining the ordering of the unscreened candidates is adapted to a carefully designed filtration. 
This filtration controls the information available to the analyst at each step, including all previously screened samples, among other information.
As the screening proceeds, more information can be used to improve the ordering,
which in turn leads to selections that are better aligned with the user's objectives.
%
The adaptivity of ACS allows practitioners to pursue a range of goals 
that are not feasible under the static structure of CS.  We summarize some of them below. 

\begin{figure}[h!]
\centering
\begin{minipage}{0.3\textwidth}
\centering 
\includegraphics[width=\textwidth]{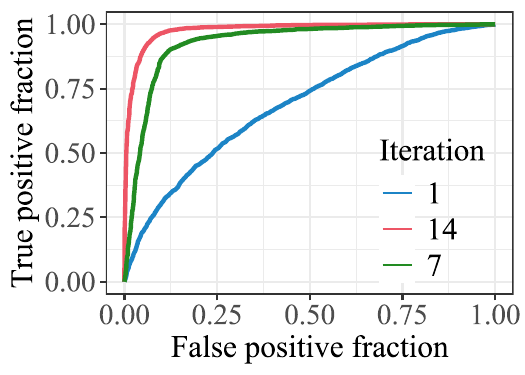}
(a) Model update
\end{minipage}
\begin{minipage}{0.3\textwidth}
\centering 
\includegraphics[width=\textwidth]{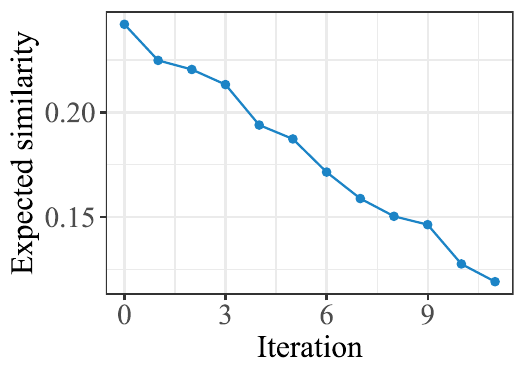}
(b) Diversified selection
\end{minipage}
\begin{minipage}{0.3\textwidth}
\centering 
\includegraphics[width=\textwidth]{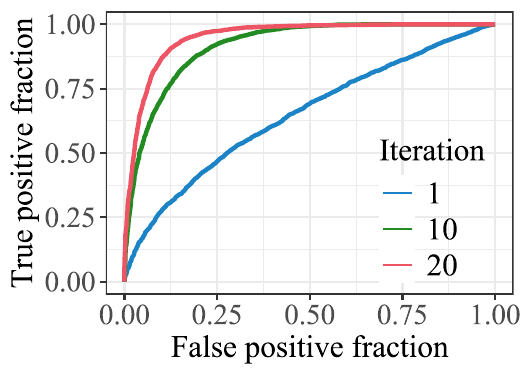}
(c) New labels
\end{minipage}
\caption{(a) The ROC curves of the (selected) prediction model 
at the $1^{\text{st}}$, $7^{\text{th}}$, and $14^{\text{th}}$ model updates;
simulation details are provided in Section~\ref{sec:adaptive_model_selection}.
(b) The expected similarity of the unscreened units with positive labels 
versus the screening rule updates; simulation details are provided in Section~\ref{sec:diversified_selection}. (c) The ROC curves of the fitted random 
forest classifier at the $1^{\text{st}}$, $10^{\text{th}}$, 
and $20^{\text{th}}$ model updates; simulation details are in Section~\ref{sec:newlabels_sim}.
}
\label{fig:ensemble_roc}
\end{figure}

\vspace{0.5em}
\begin{itemize}
\item {\bf Model update and selection:} 
As more calibration data are screened, these labeled data can be incorporated to train a new model or conduct model selection. 
This improves the model's ability to correctly order the remaining candidates---by distinguishing promising from non-promising ones,
which ultimately leads to higher selection power. We demonstrate this 
phenomenon in panel (a) of Figure~\ref{fig:ensemble_roc}: in a simulation study, 
we track the fitted model's {\em receiver operating characteristic (ROC)} curve 
after the 1st, 7th, and 14th model updates. As the metric 
improves over updates, we observe a corresponding increase in selection power, 
as shown in Section~\ref{sec:simulation}.
\item {\bf Flexible preference:} Beyond optimizing power, the adaptivity of ACS allows the analyst to prioritize different  
considerations at different stages of the screening process.
In drug discovery, a scientist may want to prioritize drug candidates with high binding affinity at the beginning, 
but as more candidates are screened, they may shift their preference toward drug candidates that are 
substantially different from the existing campaigns. 
With ACS, the analyst is free to update the scoring function---whether subjective or data-driven---to reflect evolving preferences, such as increasing the diversity of selected candidates.
Panel (b) of Figure~\ref{fig:ensemble_roc} plots a similarity measure among the remaining units over iterations. The decreasing similarity measure indicates increasing diversity.

\item {\bf Incorporating new labeled data:}  
In some applications, the screened (unselected) test candidates may be sent to a human expert for labeling (e.g., in the LLM filtering example introduced earlier).
The newly labeled data can be integrated into later steps of the selection process, 
such as refining the fitted model (as demonstrated in Panel (c) of Figure~\ref{fig:ensemble_roc}). We will show in simulations and 
real-data applications in LLM deployment that this incorporation of 
new labels leads to improved selection power. 
\end{itemize}

\subsection{Organization of the paper}
The rest of the paper is organized as follows.
In Section~\ref{sec:setup}, we introduce the problem setup and the notation used throughout the paper.
We present the general framework of ACS in Section~\ref{sec:method} and its 
instantiations 
in Section~\ref{sec:generalization} for several specific desiderata introduced earlier.
Section~\ref{sec:simulation} presents simulations that demonstrate the performance of ACS;
Sections~\ref{sec:real_data} and~\ref{sec:drug_discovery} apply ACS to real-world applications in LLM deployment and drug discovery, respectively.
The paper is concluded with a discussion in Section~\ref{sec:discussion}.

\section{Preliminaries}
\label{sec:setup}

\subsection{Problem setup}
Let $X \in \cX$ denote the feature vector, and $Y \in \RR$ the outcome of interest.
Consider a set of labeled data $\cD_\lab = \{(X_i,Y_i)\}_{i=1}^n$, 
and  test set $\cD_\te = \{X_{n+j}\}_{j=1}^m$, 
whose outcomes $\{Y_{n+j}\}_{j=1}^m$ are unobserved. 
We also denote $Z_i = (X_i,Y_i)    $ for $i \in [n+m] := \{1,\ldots,n+m\}$ for notational convenience.   
For each unit $i\in[n+m]$, the property of interest is characterized by a set $\cC_{i}$; 
we wish to pick out the test units with $Y_{n+j} \notin \cC_{n+j}$. 
For example, when the goal is to detect the samples with $Y_{n+j} > 0$,
we set $\cC_{n+j} = (-\infty,0]$. 
The only assumption we make on the data-generating process is as follows.

\vspace{0.5em}
\begin{assumption}
\label{assumption:exchangeable}
The property set $\cC_i$ is observed for each $i\in[n+m]$, and  
$\{(X_i,Y_i,\cC_i)\}_{i \in [n+m]}$ are jointly exchangeable across $i\in[n+m]$.
\end{assumption}
\vspace{0.5em}

Assumption~\ref{assumption:exchangeable} is quite mild and can be satisfied in many practical scenarios, 
for example, when the $Z_i$'s are i.i.d.~samples from some distribution, 
and the property sets $\cC_i \equiv \cC$ for some deterministic set $\cC$
or $\cC_i = c(X_i)$  for some deterministic
function $c: \cX \mapsto \RR$. Following~\citet{jin2023selection}, we approach this problem from a testing perspective:
for each $j\in[m]$, consider the null hypothesis 
\$
H_{0,j}: Y_{n+j} \in \cC_{n+j}.
\$ 
When $H_{0,j}$ is rejected, we declare the $j^{\text{th}}$ test sample to be interesting,
and the selection problem can now be viewed as a multiple testing problem.
We aim to design a selection algorithm $\cA$ that takes as input $\cD_\lab$ and $\cD_\te$
and outputs a set of indices $\cS \subseteq [m]$ that are declared as interesting.
The FDR and power of $\cS$ are defined as follows: 
\@ 
&\fdr(\cS) = \EE[\fdp(\cS)], \qquad 
\fdp(\cS) = \frac{\sum_{j=1}^m \ind\{j \in \cS, Y_{n+j} \in \cC_{n+j}\}}{|\cS|},\\
&\textnormal{Power}(\cS) = \EE\bigg[\frac{\sum_{j=1}^m \ind\{j\in\cS, Y_{n+j} \notin \cC_{n+j}\}}{\sum^m_{j=1} \ind\{Y_{n+j} \notin \cC_{n+j}\}}\bigg], \label{eq:def_power}
\@
where we follow the convention that $0/0 = 0$ throughout the paper,
and the expectation is taken over the randomness in $\cD_\lab$ and $\cD_\te$.
We will use $\cH_0 = \{j \in [m]: Y_{n+j} \in \cC_{n+j}\}$ hereafter 
to denote the set of true null hypotheses (i.e., ``uninteresting'' samples),
noting that $\cH_0$ is random. Given a pre-specified FDR level $\alpha \in (0,1)$, 
the goal is to select as many test samples as possible that are 
``interesting''  while controlling the FDR at level $\alpha$.

\subsection{Background: conformal selection}
\label{subsec:cs}

The problem described above was first considered in~\citet{jin2023selection}, with $\cC_i$ in the form of $(-\infty,c_i]$
for some threshold $c_i \in \RR$. They proposed the conformal selection (CS) procedure that  
builds upon the split conformal inference framework~\citep{vovk2005algorithmic,papadopoulos2002inductive,lei2015conformal}. 
The CS procedure starts by splitting the training set $\cD_\lab$ into two parts:
a model-fitting set $\cD_\tr$ and a calibration set $\cD_\ca$. 
With slight abuse of the notation, we use $\cD_\tr$, 
$\cD_\ca$ and 
$\cD_\te$ to denote either the samples or their indices 
when no confusion can arise from the context.
Using $\cD_\tr$, a predictive model $\hat{f}: \cX \mapsto \RR$ is fitted; 
a conformal $p$-value~\citep{vovk2005algorithmic,bates2023testing} is then computed for each test sample $j\in[m]$
by contrasting its predicted value $\hat f(X_{n+j})$ with the predictions on the calibration set $\cD_\ca$:
\begin{equation}
\label{eq:conf_pval}
p_j = \frac{1+\sum_{i \in \cD_\ca} \ind\big\{\hat f(X_{n+j}) \ge  \hat f (X_i) , Y_i \le c_i \big\}}{|\cD_\ca| + 1}.
\end{equation} 
It can be proved that the conformal p-values are valid p-values,
i.e., $\PP(p_j \le t, Y_{n+j}\le c_{n+j}) \le t$ for any $t\in[0,1]$;\footnote{There is 
another version of the conformal p-value involving a tie-breaking random variable $U_j\sim \text{Unif}[0,1]$, 
which is not introduced here for simplicity. See~\citet{jin2023selection} for details and~\citet{gui2024conformal} for the connection.}
meanwhile, a small $p_j$ suggests evidence against the null hypothesis $H_{0,j}$, 
indicating that the $j^{\text{th}}$ test sample is likely to be interesting.
The CS procedure finally selects the test samples by applying the Benjamini-Hochberg (BH) procedure~\citep{benjamini1995controlling}
to the conformal p-values at level $\alpha$.~\citet{jin2023selection} show that CS controls the FDR at level $\alpha$ under 
Assumption~\ref{assumption:exchangeable}.

As mentioned in the introduction, CS is a one-shot procedure, requiring the analysis plan---e.g., the choice 
of prediction models, the type of scoring functions and the data usage---to be 
pre-determined before inspecting the data. 
Our work generalizes CS to enable adaptive data analysis while maintaining the FDR control.
In our new proposal, one can iteratively improve the prediction model and perform data-driven model selection 
to boost the power of the selection procedure, and/or update the utility function based on (part of) the observed data adaptively.

\subsection{Related work}
This work relates to the line of research on model-free selection~\citep{jin2023selection},
with applications in drug discovery~\citep{jin2023model,bai2024conformal} and LLM deployment~\citep{gui2024conformal}. 
Recently,~\citet{bai2024optimized} extended CS to
perform model selection in a data-driven manner aimed at power improvement.
While this aligns with one of our key goals,
the adaptivity of ACS offers additional flexibility: it enables the 
incorporation of expert preferences, supports complex objectives like diversity, 
and allows integration of new labeled data.
Moreover, the techniques used in~\citet{bai2024optimized} are substantially different from ours.
Other works have extended the model-free selection problem to various settings:
(1) settings beyond the exchangeable data assumption~\citep{jin2023model,lee2025selection},
(2) online selection~\citep{xu2024online,huo2024real}, 
and (3) settings where the goal is to optimize metrics beyond the size of selection subject to constraints~\citep{huo2024real,wu2024optimal,nair2025diversifying}. The instantiation of ACS to promote diversity is most closely related to 
the third category---particularly~\cite{wu2024optimal}, 
who asymptotically optimize diversity metrics of the selection set, 
and~\cite{nair2025diversifying}, who leveraged a similar martingale structure to
ours that prunes the conformal selection set for enhanced diversity. 
In contrast to these approaches, ACS pursues diversity by adaptively
updating the priority of unscreened candidates, rather than performing 
a global optimization over the selection set. Unlike~\citet{wu2024optimal},
ACS provides finite-sample FDR control without requiring any modeling assumptions.
Compared with~\cite{nair2025diversifying}, who always returns a subset of the original conformal selection set, ACS can---in some cases---improve both the diversity and the number of selected candidates by iteratively reordering candidates
during the selection process.


Related to conformal selection, the literature on novelty detection with 
conformal p-values~\citep{bates2023testing,mary2022semi,marandon2024adaptive,bashari2024derandomized,liang2024integrative,lee2025full,bashari2025robust, lee2025finding} 
concerns picking out from a pool of samples the outliers whose 
distribution differs from that of a reference set. 
While their setting is distinct from ours, our proposed method is also applicable to the novelty detection problem with slight modification. 

Another related line of work is adaptive data analysis with statistical validity~\citep{dwork2015preserving,tian2018selective,lei2018adapt,weinstein2017power,lei2021general,balakrishnan2019interactive,duan2020familywise,yang2021bonus,chao2021adapt,ren2023knockoffs,roy2025exploration},   
where the research question is to design a protocol for human-in-the-loop data analysis  
that limits the amount of information accessible to the data analyst at different stages, 
so as to ensure the statistical validity in the presence of double-dipping. 
Our approach is similar in spirit, where the adaptivity and the statistical validity are both achieved by carefully controlling the information 
flow in the selection procedure. In particular, our design of protocol is inspired by that of~\citet{yang2021bonus};  
they, however, focus on testing parametric hypotheses, while our hypotheses of interest are fully nonparametric.

\section{Adaptive conformal selection}
\label{sec:method}

\subsection{Warm-up: CS as a screening procedure}

As a warm up, 
we first give an alternative interpretation of CS 
which builds the key intuition for ACS. 
To streamline the exposition, we set $\cC_i \equiv (-\infty,0]$ in this part. 

\begin{figure}[ht]
\centering 
\includegraphics[width=0.7\textwidth]{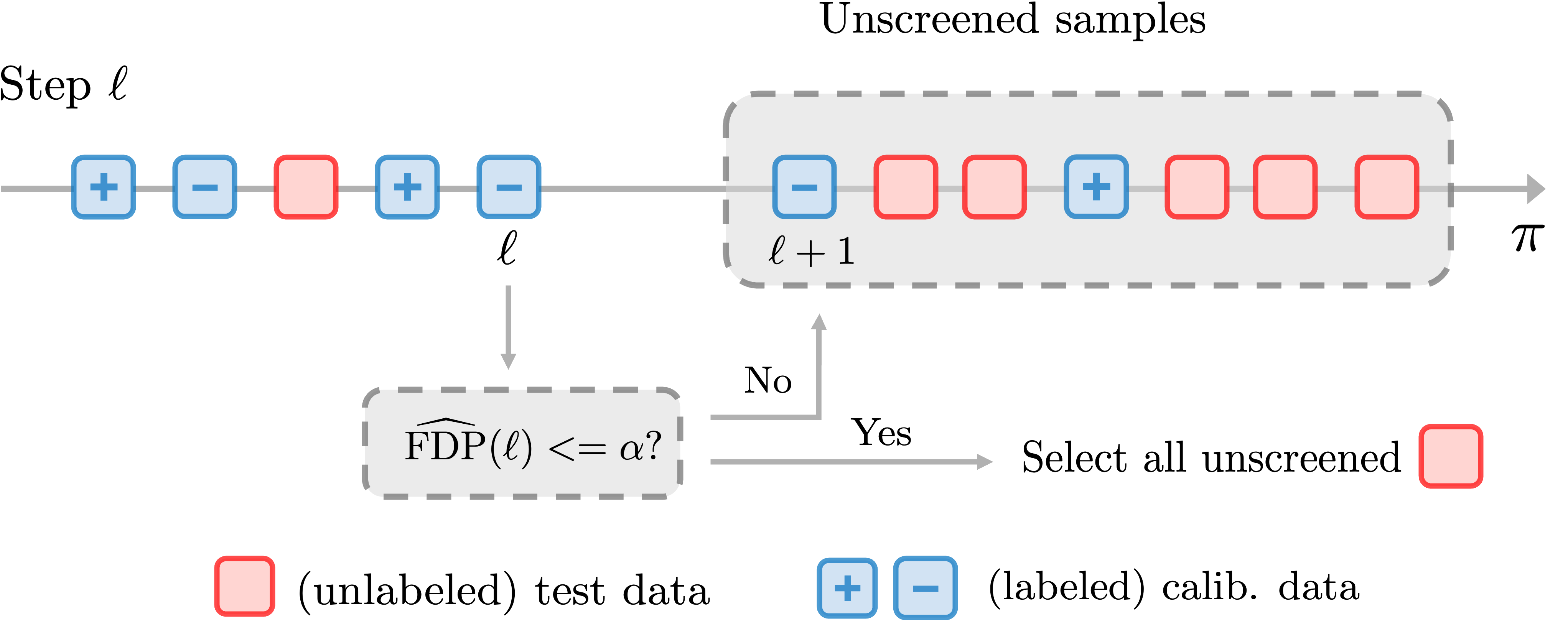}    
\caption{An illustration of the CS procedure as a sequential screening process,
where the plus (resp.~minus) signs correspond to interesting (resp.~uninteresting) samples.
The goal is to select positive test samples, and 
the calibration and test samples are ordered by their predicted values $\hat{f}(X_i)$.  
At step $\ell$, CS screens the sample corresponding to $\pi(\ell)$ if $\hat{\fdp}(\ell)$ is above the target level $\alpha$;
otherwise it stops the screening and selects all the unscreened test samples.}
\label{fig:cs}
\end{figure}

In addition to the p-value perspective from~\cite{jin2023selection} introduced in Section~\ref{subsec:cs}, 
CS can also be interpreted as a sequential screening procedure~\citep{mary2022semi}. 
Specifically, after fitting the prediction 
model on $\cD_\tr$, each of the calibration and test samples is assigned a predicted value $\hat f(X_i)$. 
CS then ranks the samples in ascending order of $\hat f(X_i)$, 
from least to most promising. 
For notational convenience, let $k = |\cD_\tr|$, and  let $\pi$ 
be a permutation of $\{k+1,\ldots,n+m\}$ such that
$\hat{f}(X_{\pi(k+1)}) \le \ldots \le \hat{f}(X_{\pi(n+m)})$.
CS proceeds by sequentially screening the samples in the order specified by $\pi$. 
At each step $\ell\geq k+1$, 
it computes the FDP estimate for the set of candidates ranked above $\pi(\ell)$, namely, 
\$
\hat{\fdp}(\ell) = \frac{m}{n-k+1} \cdot \frac{1 + \#\{j > \ell: \pi(j) \text{ is from }\cD_\ca \text{ and } Y_{\pi(j)}\le 0 \}}
{\#\{j > \ell: \pi(j) \text{ is from }\cD_\te\}}. 
\$
The screening stops the first time $\hat{\fdp}(\ell) \le \alpha$, 
and the remaining test samples are declared as interesting---this is the selection set $\cS_{\text{CS}}$.
See Figure~\ref{fig:cs} for an illustration of the screening process.

From this sequential screening perspective, the ordering $\pi$ controls two aspects of the selection process: (1) the form of the final selection set, i.e., the relative ordering of the candidates, and (2) the stopping time---i.e., how many top-ranked candidates are selected---determined by the estimated FDP at each step.  
Intuitively, a scientist seeking diversity among selected drug candidates should choose an ordering $\pi$ that avoids clustering similar drugs in nearby ranks to reduce the risk of selecting a homogeneous set. 
A practitioner seeking high power should aim for an ordering $\pi$ in which the test and positive calibration samples appear later in the screening  (i.e., highly ranked in $\pi$) to decrease the FDP estimate.   

To achieve these goals, it is highly desirable to adjust the ordering $\pi$ after inspecting some of the calibration and test data. While such behavior is prevented in CS,  we will show that ACS allows to update $\pi(\ell),\dots,\pi(n+m)$ at every step $\ell$ using a rich set of information without compromising the FDR control, thereby enhancing the flexibility and efficacy of the selection procedure. 

\subsection{ACS: the general framework}
As in CS, ACS sequentially screens the samples from the least promising to 
the most promising until a stopping criterion is met, then declaring the unscreened 
test samples as interesting. 
Importantly, the ordering of the unscreened samples can be updated adaptively with 
the newly available information revealed at each step. 
We formally describe the general procedure of ACS below.

\paragraph{Initialization.}
Before running ACS, we first randomly permute the order of all samples. Without loss of generality, let $[n+m]$ denote the indices of permuted samples, and $i\in [n+m]$ provides no information about whether the sample $i$ belongs to the labeled set or not. 
For each $i\in[n+m]$,  
we define an indicator $A_i$ such that $A_i = 0$ (resp.~$A_i = 1$) if $i$ corresponds to a labeled (resp. test) sample; we refer to $A_i$ as the ``membership information''.  
Due to the permutation, one cannot tell whether $X_i$ is from the labeled or 
test set without observing $A_i$. 
The goal of \algoname~is to adaptively determine a permutation $\pi$ of $[n+m]$ that specifies the screening order, where the sample to be screened at step $\ell$ is $\pi(\ell)$. 
One may reserve a subset of size $k\geq 0$ among the labeled samples 
to initialize model training. Without loss of generality, 
we take $(\pi(1),\dots,\pi(k))$ as a random subset of $\{i\colon A_i=0\}$. 

\paragraph{Adaptive screening.} 
Next, for any $\ell \ge k$,
we are to decide the next-to-screen sample $\pi(\ell+1)$ from the set of as-yet unscreened indices at time $\ell$. 
We define the following quantities that will be useful for introducing the algorithm:
\begin{itemize}
\item The indices of already screened samples: $O_\ell := (\pi(1),\ldots,\pi(\ell))$;  
\item The index sets of non-null and null unscreened labeled samples: 
\$
N^+_\ell = \{i \in[n+m]\backslash O_\ell: A_{i} = 0, Y_{i} \not\in \cC_{i}\},~ 
N^-_\ell= \{i \in [n+m] \backslash O_\ell: A_{i} = 0, Y_{i} \in \cC_{i}\};
\$ 
\item The set of indices of unscreened test samples:
$P_\ell = \{i \in [n+m] \backslash O_\ell \colon A_i = 1\}$;
\item 
The index set of all the  unscreened samples: 
$U_\ell = [n+m] \backslash O_\ell = N^+_\ell \cup N^-_\ell \cup P_\ell $.
\end{itemize} 
The only requirement that we impose on the screening order is that $\pi(\ell)$ is measurable with respect to 
the $\sigma$-algebra defined as follows:
\@\label{eq:filtration_alg}
\cF_{\ell} = \sigma\left(O_\ell, |N^-_{\ell}|, |P_\ell|, 
\{(\tZ_i,A_i)\}_{i \in O_\ell \cup N^+_k}, \{X_i\}_{i\in U_\ell} \right). 
\@ 
Above, $|A|$ denotes the cardinality of a set $A$; $\tZ_i = Z_i$ if $A_i = 0$ and $\tZ_i = X_i$ otherwise. Note that in CS, $\{\pi(\ell)\}_{\ell > k}$ is determined purely by $\{\tilde Z_i, A_i\}_{i \in O_k}$ and $\{X_i\}_{i\in U_k}$, and therefore 
CS can be viewed as a special case of ACS.

The key restrictions imposed by $\cF_\ell$ are as follows. 
First, the full data and membership information are revealed for $O_\ell\cup N_k^+$---that is, 
the previously screened (both labeled and test) samples and all the non-null labeled samples; 
these data can be used for training purposes in several applications of \algoname, 
especially the labeled samples in them. 
Second, the term $\{X_i\}_{i\in U_\ell}$ means the covariates for the unscreened samples can be used without revealing their membership information; even though we do not know whether they belong to the test set, this can still be useful in several cases, e.g., checking the characteristics---such as diversity---of the unscreened candidates to gauge the screening preference. 
Finally, the only information we could use for the unscreened null labeled and test samples is their cardinality, i.e., $|N_\ell^-|$ and $|P_\ell|$. 
These constraints specify precisely the information available to the user in $\cF_\ell$.
We will see that this flexible and adaptive use of the sequentially revealed information in $\cF_\ell$ enables us to use \algoname~for various distinct goals, described the next section.

\paragraph{Stopping rule.} Finally, we specify how \algoname~stops the screening process and returns a final selection set. 
At each step $\ell \ge k$, we define an FDP estimate among the remaining unscreened test samples as 
\@\label{eq:fdp_est}
\widehat{\fdp}(\ell) = \frac{m}{n-k+1} \cdot \frac{1+ |N^-_{\ell}|}{|P_{\ell}| \vee 1}, 
\@ 
where $a \vee b = \max(a,b)$.
Intuitively, under the screening rule, the labeled and test null samples are exchangeable, so we would expect (roughly): 
\$ 
\frac{1+ |N^-_\ell|}{n-k+1} \approx \frac{|P_\ell \cap \cH_0|}{m},
\$ 
which renders $\widehat{\fdp}(\ell)$ a reasonable estimate of the true FDP among the unscreened test samples. 
Whenever $\hat{\fdp}(\ell) \le \alpha$, we stop the screening process and declare 
the remaining test samples $P_\ell$ as selection. Otherwise, we continue to the next step and determine $\pi(\ell+1)$
using the above principle.\\ 

\begin{figure}[htt]
\centering
\includegraphics[width=0.82\textwidth]{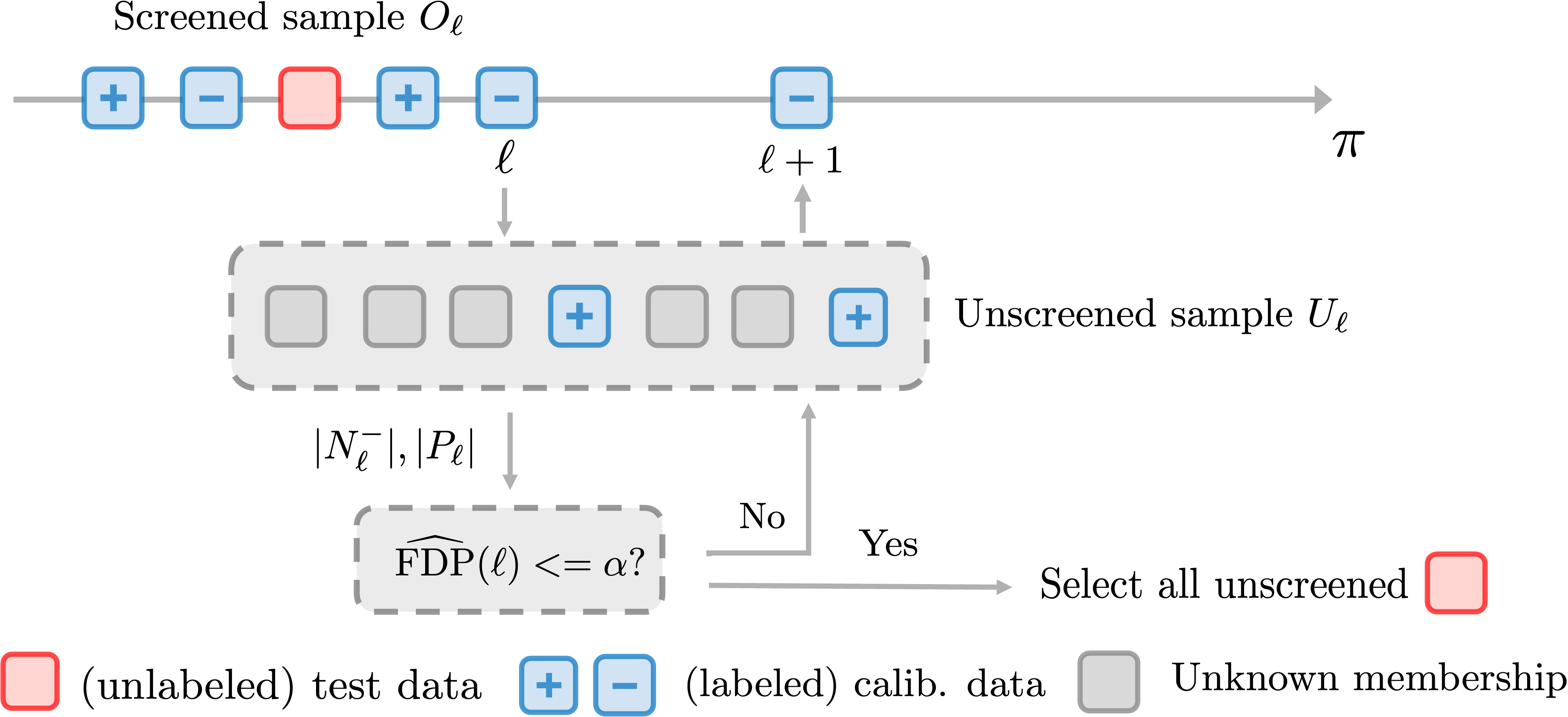}
\caption{An illustration of the ACS procedure, where the 
plus (resp.~minus) signs correspond to the interesting (resp.~uninteresting) samples.
The goal is to select interesting test samples. 
At step $\ell$, if the FDP estimate $\widehat{\fdp}(\ell)$ is above the target level $\alpha$, 
ACS determines the next sample to screen $\pi(\ell+1)$ from the unscreened candidates $U_\ell$ using the filtration $\cF_\ell$ defined in~\eqref{eq:filtration_alg}; 
otherwise, it stops the screening and declares all the unscreened test samples as selection.}
\label{fig:acs}
\end{figure}

To summarize, 
ACS first assigns $\{\pi(i)\}_{i=1}^k$ as a random subset of those $A_i=0$ for initialization; it then sequentially determines $\pi(\ell)$ for $\ell \ge k+1$,
requiring that $\pi(\ell)$ is measurable with respect to 
$\cF_{\ell-1}$. It returns the final selection set $\cS_{\algoname} = P_T$, 
where we define the stopping time
\@\label{eq:def_T}
T = \inf\{\ell \ge k: \widehat{\fdp}(\ell) \le \alpha \},
\@ 
with the FDP estimator defined in~\eqref{eq:fdp_est}. 
The complete algorithm is summarized in Algorithm~\ref{alg:acs} and a graphical illustration is provided in Figure~\ref{fig:acs}.

\begin{algorithm}[htbp]
    \caption{ACS: adaptive conformal selection}
    \label{alg:acs}
    \begin{algorithmic}[1]
    \Require Labeled samples $\{(X_i,Y_i)\}_{i \in [n]}$; 
    test samples $\{X_{n+j}\}_{j \in [m]}$; target FDR level $\alpha$;
    initial training sample size $k$. 
    \vspace{0.5 em}
    \State Randomly permute the indices of all samples.
    \State Initialize the screening order $\pi(i)$ for $i \in [k]$ by randomly choosing $k$ samples with $A_i = 0$. 
    \For{$\ell = k,\ldots, n+m$} 
    \State Compute FDP estimate $\widehat{\fdp}(\ell)$ using \eqref{eq:fdp_est}. 
    \State \If{$\widehat{\fdp}(\ell) \le \alpha$}
    \State \textbf{break} \Comment{Stop screening and declare the remaining test samples as interesting.}
    \EndIf
    \State Determine  $\pi(\ell+1) \in U_\ell$ using $\cF_{\ell}$ given in~\eqref{eq:filtration_alg}.
    \EndFor 
    \vspace{0.5 em}
    \Ensure Rejection set $P_\ell^+$. 
    \end{algorithmic}
    \end{algorithm}

\subsection{Theoretical guarantees}
The following theorem establishes the FDR control of~\algoname.

\vspace{0.5em}
\begin{theorem}\label{thm:fdr_control}
Under Assumption~\ref{assumption:exchangeable}, for any fixed $\alpha\in(0,1)$, the output of~\algoname~applied at level $\alpha$ controls the FDR below $\alpha$.  
\end{theorem}
\vspace{0.5em}
We provide a sketch of the proof of Theorem~\ref{thm:fdr_control} and 
defer the details to Appendix~\ref{appd:proof_fdr_control}. 
First, by the choice of $T$ and the definition of the FDP estimator, 
\$
\fdr(\cS_{\algoname}) = \EE\bigg[\frac{|P_T \cap \cH_0|}{|P_T|}\bigg] 
= \EE\bigg[\widehat{\text{FDP}}(T) \cdot 
\frac{n-k+1}{m} \frac{|P_T\cap \cH_0|}{1 + |N_T^-|}\bigg] 
\le \alpha \frac{n-k+1}{m} \EE\bigg[\frac{|P_T\cap \cH_0|}{1 + |N_T^-|}\bigg].
\$
At each step $\ell \ge k$, the elements in $P_\ell \cup N_\ell^-$ remain exchangeable despite the adaptive data analysis, since the update rule of ACS makes no use of their membership information. Using this fact, we prove that  
\$
M_\ell := \frac{|P_\ell\cap \cH_0|}{1+ |N_\ell^-|}
\$
is a super-martingale with respect to a filtration that is slightly richer than $\{\cF_\ell\}_{\ell \ge k}$.
The proof is completed by noting that $T$ is a stopping time with respect to the same filtration and 
applying the optional stopping theorem.
A similar martingale structure is used in~\cite{nair2025diversifying} to construct e-values for optimizing certain objectives---such as diversity---of the selected samples in conformal selection. However, the specific techniques differ significantly in the two works. 


\vspace{0.5em}
\begin{remark}
As discussed earlier, ACS strictly generalizes the original framework of CS~\citep{jin2023selection}:
CS uses $\{Z_i\}_{i \in O_k}$ to train a prediction model 
$\hat{g}$, which outputs scores measuring 
the likelihood of samples being interesting (non-null)
and then uses the scores to determine the screening order. 
That is, $\{\pi(\ell)\}_{\ell >k }$ is purely determined by $\{(\tilde Z_i, A_i)\}_{i\in O_k}$
and $\{X_i\}_{i\in U_k}$, which is a subset of $\cF_\ell$, for $\ell > k$.

\end{remark}
\vspace{0.5em}

\begin{remark}[Connection to existing adaptive multiple testing methods]
Our interactive testing framework is closely related to the broader field of 
adaptive multiple testing. In particular,~\citet{lei2018adapt,yang2021bonus} 
propose procedures that sequentially use partially revealed data to update rejection thresholds 
for each hypothesis, stopping once the estimated FDR falls below the 
target level. In both cases, the FDR control is achieved by constructing 
suitable martingales. As noted before,~\citet{yang2021bonus} focus 
on testing parametric null hypotheses, while we consider fully nonparametric and random hypotheses. 
While the underlying principle is similar, 
our goals and implementation of this general idea  
differ significantly from those of prior works.  

\end{remark}

\section{Instantiation of ACS}
\label{sec:generalization}
In this section, we present three instantiation of~\algoname, each serving distinct goals.  
Section~\ref{sec:power} focuses on power improvement with adaptive model update and selection,
Section~\ref{sec:diversity} on diversifying the selection set, 
and  Section~\ref{sec:newlabels} is about incorporating newly available labels for the 
screened-out units.

\subsection{Boosting power}
\label{sec:power}
The screening order $\pi$ plays a crucial role in the power of ACS.
Intuitively, 
we would like to screen the least promising samples---those unlikely to be non-null test units---first, 
leaving the more promising 
ones towards the end: in this way, the FDP estimator stays lower and the sequential procedure has a higher chance of stopping 
earlier and selecting more interesting units. 
Naturally, this necessitates an accurate prediction model to discern null and non-null samples. 

\paragraph{Model re-fitting.} As the first instantiation of ACS, we propose to re-fit the prediction  
model periodically as more labeled samples enter the filtration $\cF_\ell$. 
More specifically, at step $\ell \ge k$, we construct  
$\hat g_\ell: \cX \mapsto \RR$ via 
\$
\widehat g_\ell = \cG(Z_{\pi(1)},\ldots,Z_{\pi(k)},\tZ_{\pi(k+1)},\ldots,\tZ_{\pi(\ell)},
X_{\pi(\ell+1)},\ldots,X_{\pi(n+m)}),
\$
recalling that $\tZ_i = Z_i$ if $A_i = 0$ and $\tZ_i = X_i$ otherwise; 
$\cG$ is a model-fitting algorithm outputting a function that predicts 
the likelihood of a sample being interesting. 
The next-to-screen index $\pi_{\ell+1}$ is then determined by
\$ 
\pi_{\ell+1} = \argmin{j \in U_\ell}~\widehat g_\ell(X_j).
\$
It is clear that $\pi_{\ell+1}$ is adapted to $\cF_{\ell}$, and therefore  this instantiation of ACS controls the FDR by 
Theorem~\ref{thm:fdr_control}. 
We note that in theory the \emph{unscreened} non-null samples $\{Z_i\}_{i \in N_\ell^+}$ can also be used to update the model,
i.e., allowing $\cG$ to depend on $\{Z_i\}_{i \in N_\ell^+}$. However, we do not pursue this strategy here, as empirically we do not observe it to significantly improve the power.

The model-fitting algorithm $\cG$ can, in principle, be any machine learning algorithm.
\citet{jin2023selection} shows that the choice of score maximizing the 
asymptotic power under FDR constraints should be a monotone function of  
$\PP(Y \notin \cC \given X)$, i.e., 
the conditional probability of a sample being interesting given the covariates. 
Guided by this result, 
we suggest fitting a classification model for $\ind\{Y \notin \cC\}$ over $X$ 
on the data $\{Z_i\}_{i = 1}^k \cup \{Z_{\pi(i)}: A_{\pi(i)}=0, k+1\le i \le \ell\}$, i.e., 
all the labeled samples that have been screened at step $\ell$; 
then, we set $\hat g_\ell(x)$ to be the predicted probability of $Y \notin \cC$ for $x$. 
As the screening procedure proceeds, more calibration samples can be used 
to fit the model, thereby improving its prediction accuracy (as visualized in
panel (a) of Figure~\ref{fig:ensemble_roc} in 
the introduction), 
the ordering of the remaining 
units, and ultimately the power of the selection procedure. 

\paragraph{Adaptive model selection.}
Our adaptive framework also supports data-driven model selection 
while preserving FDR control. 
Suppose we have candidate models $\cG_1,\ldots,\cG_H$. 
At step $\ell \ge k$, we may want to pick the best model among the candidates 
to guide the subsequent screening order. We propose a 
cross-validation-based procedure for model selection using the available information. 

Let $\cD_{\lab,\ell}:= 
\{Z_i\}_{i=1}^k \cup \{Z_{\pi(i)}: A_{\pi(i)}=0, k+1\le i\le \ell\}$ 
denote all the screened labeled samples available at timestep $\ell$. 
We randomly split $\cD_{\lab,\ell}$ into $K$ non-overlapping folds $\{\cD^{(k)}_{\lab,\ell}\}_{k=1}^K$. Then, 
for each $k \in [K]$ and each $h \in [H]$, we obtain a selection set $\cS_{h,k}$ by running the CS procedure with 
prediction model $\cG_h$, treating  
$\cD_{\lab, \ell}^{(-k)}:= \cD_{\lab,\ell} \backslash \cD_{\lab,\ell}^{(k)}$ as the training set, $\cD_{\lab,\ell}^{(k)}$ 
as the calibration set, and $\{X_i\}_{i\in U_\ell}$ as the test set. We then measure the performance 
of $\cG_h$ by the average number of selections across the $K$ folds, and select the best model as 
\$
\hat{h}_\ell = \textstyle{\argmax{h \in [H]}~\frac{1}{K}\sum_{k\in [K]} }|\cS_{h,k}|.
\$
Since CS is agnostic to the order of the test samples, 
one can check that the proposed model selection procedure is adapted to $\cF_\ell$, 
thereby preserving FDR control by Theorem~\ref{thm:fdr_control}.

\paragraph{Model update frequency.}
In principle, the prediction model can be updated at every step as new information in $\cF_\ell$ becomes available.
However, this may be computationally expensive in practice.
To balance computational cost and power improvement, one may the model every $L$ steps, 
where the hyperparameter $L$ can be chosen by the user based on their computational constraints. 

\subsection{Diversifying selection}
\label{sec:diversity}
In many applications, in addition to maximizing the power, 
it is desired to optimize the diversity of the selection set. In drug discovery, 
scientists may want to include drug candidates with distinct action mechanisms in the selection set. 
However, with a pre-specified score function, CS is highly likely to select many similar drugs with similar predicted values. Since the diversity depends on the composition of the entire selection set, diversifying conformal selection while preserving FDR control proves challenging. Existing works either resort to asymptotic analysis assuming accurate modeling~\citep{huo2024real} or require complex recursive optimization~\citep{nair2025diversifying}. 

Our second instantiation presents a novel solution to this problem by dynamically promoting ``more diverse'' drugs based on available information in the process. The intuition is that as more drugs are screened, the information in $\cF_\ell$ allows the analyst to inspect the diversity of the unscreened drugs and design strategies to promote less-visited ones.  
For instance, if many test drugs in one region have been screened (i.e., excluded from the selection set), 
the remaining drugs in this region may be favored in subsequent rankings, so that they have a higher chance to be selected. 

Formally, we introduce a function $\theta: \cX \times \cX \mapsto \RR$ 
to measure the similarity between two units.
Common choices for $\theta$ include the {\em radial basis function (RBF)} kernel 
$\theta(x_i, x_j) = \exp(-\|x_i - x_j\|^2/\sigma_0^2)$ for some user-specified 
$\sigma_0 > 0$, or the {\em cosine similarity} $\theta(x_i, x_j) = \frac{x_i^\top x_j}{\|x_i\| \|x_j\|}$; 
see~\citet{yang2015multi,zhong2017learning,wu2024optimal,huo2024real} for more details. Given $\theta$, 
the diversity of a set of units $\cS \subseteq [m]$ can be measured by the {\em expected similarity (ES)}:
\$ 
\text{ES}_{\theta}(\cS)  =  
\EE\bigg[\frac{\sum_{1 \le i < j\le m}\ind\{i \in \cS,j\in \cS\}\delta_i\delta_j \cdot \theta(X_{n+i},X_{n+j})}
{\sum_{1 \le i < j\le m}\ind\{i\in \cS, j\in \cS\}\delta_i\delta_j } \bigg],
\$
where $\delta_i = \ind\{Y_{n+i} \notin \cC_{n+i}\}$, for $i\in [m]$. 
By definition, $\text{ES}_{\theta}(\cS)$ measures the similarity of the correctly selected units in $\cS$,
and is smaller when the selected units are more diverse. 
Roughly speaking, we aim to optimize a weighted combination of the power and diversity of a selection set $\cS\subseteq [m]$: 
\$
L(\cS;\lambda) = (1-\lambda) \cdot \text{Power}(\cS) - \lambda \cdot \text{ES}_{\theta}(\cS), 
\$ 
where $\lambda \in [0,1]$ is a user-specified parameter that balances the two objectives. When $\lambda = 0$, we recover the original power-based selection, 
and when $\lambda = 1$, the selection is solely based on diversity.

\subsubsection{Diversity-aware ordering}\label{sec:diversity-aware-ordering}

In order to connect the ordering strategy in~\algoname~with the objective function $L(\cS;\lambda)$,  
our approach produces a score for each test unit, which will be used in the ordering~\citep{xie2024boosted}. To be specific, at step $\ell \ge k$, we aim to select from 
a pool of $\tilde n := n+m - \ell$ test units given their features. To simplify the notation, we use 
$X_1,\ldots, X_{\tilde n}$ to denote these test features in this section, when no confusion can arise given the context. 

We assign each sample $i$ a score 
of the form $(1-\lambda) \hat\delta_i + \lambda \hat\xi_i$, 
where $\hat\delta_i$ promotes high power, 
and $\hat\xi_i$ promotes greater diversity. 
While the specific choice of $\hat \delta_i$ and $\hat\xi_i$ are flexible, 
we introduce below a general principle for building effective scores.
Specifically, we set $\hat \delta_i$ as the current estimate of $\PP(Y_i \notin \cC_i \given X_i)$, $\forall i \in [\tilde n]$, obtained with the available information at step $\ell$. 
For the diversity part, we use a function $\xi: \cX \mapsto [0,1]$ to represent the ``probability'' of being selected, which will be optimized for diversity subject to FDR control. Consider the population objective with 
FDR control constraints: 
\begin{equation}
\begin{aligned}
\label{eq:diversity_opt}
\min_{\xi: \cX \mapsto [0,1]}~ &\frac{\EE\big[\xi(X_1) \xi(X_2) \delta_1\delta_2 \cdot \theta(X_1,X_2)\big]}
{\EE\big[\xi(X_1) \xi(X_2) \delta_1 \delta_2\big]}\\
\text{s.t.}~&\frac{\EE\big[\xi(X_1) (1-\delta_1)\big]}{\EE[\xi(X_1)]} \le \alpha,
\end{aligned}
\end{equation}
where $Z_i = (X_i,Y_i)$ for $i\in\{1,2\}$ are i.i.d.~samples from the joint distribution of $(X,Y)$.
Intuitively, in~\eqref{eq:diversity_opt}, the objective function approximates $\text{ES}_\theta$ when the probability of selecting a unit $X_i$ is given by $\xi(X_i)$, while the constraint is a proxy for the FDR control. 
A similar relaxation strategy is adopted by~\citet{huo2024real}. 
We note that the FDR constraint in~\eqref{eq:diversity_opt} is optional, since 
the FDR control guarantee of the final selection set will always 
be guaranteed by the ACS framework regardless of the constraint.

The empirical version of the population-level problem~\eqref{eq:diversity_opt} can be solved using off-the-shelf optimization algorithms if desired. In the next, we introduce a special relaxation which admits a closed-form solution. 
To be specific, we compute  $\{\xi_i\}_{i=1}^{\tilde{n}}$ by solving
the following relaxed empirical version of~\eqref{eq:diversity_opt}: 
\begin{equation}
\begin{aligned}
\label{eq:diversity_opt_sample}
\min_{\{\xi_i\}_{i=1}^{\tilde n} }~ &\frac{\sum_{i,j=1}^{\tilde n}
\xi_i \xi_j \hat\delta_i \hat \delta_j \cdot \theta(X_i,X_j)}
{\sum_{i,j=1}^{\tilde n}\xi_i \xi_j \hat \delta_i \hat \delta_j}\\
\text{s.t.}~&\frac{\sum_{i=1}^{\tilde n} \xi_i (1-\hat \delta_i)}{\sum^{\tilde n}_{i=1}\xi_i} = \alpha,\\
& \xi^\top \hat \delta > 0. 
\end{aligned}
\end{equation}
Note that in~\eqref{eq:diversity_opt_sample}, we relax the range constraint $\xi\in [0,1]$ to $\xi^\top \hat \delta >0$, as both the objective and the constraint are invariant to 
the scaling of $\xi$ and only the relative importance of units is of interest in 
the ACS framework;
the estimated FDR is forced to be $\alpha$, since using up the budget of false discoveries only
increases the power.
Such a relaxation leads to a closed-form solution,
stated in the proposition below. 
Its proof is deferred to Appendix~\ref{appd:proof_diversity_opt}.

\begin{proposition}
\label{prop:diversity} 
Let $\Theta \in \RR^{\tilde n \times \tilde n}$ be the matrix with $(i,j)$-the entry 
$\hat \delta_i \hat \delta_j\theta(X_i,X_j)$ and let $\hat \delta = (\hat \delta_1,\ldots,\hat \delta_{\tilde n})$.
Assume that $\Theta$ is positive definite.
Define $a = \hat \delta^\top \Theta^{-1}\one_{\tilde n}$, 
$b = \hat \delta^\top \Theta^{-1} \hat \delta$, 
and $c = \one_{\tilde n}^\top \Theta^{-1} \one_{\tilde n}$.
The optimal solution to~\eqref{eq:diversity_opt_sample} is given by 
\$
\xi^*  \propto \xi_0 := \Big(\frac{b}{1-\alpha}-a\Big) \Theta^{-1}\one_{\tilde n} - 
\Big(\frac{a}{1-\alpha}-c\Big)\Theta^{-1}\hat \delta. 
\$
\end{proposition}

In our implementation, we threshold $\xi_0$ at $0$ and normalize 
it to be bounded in $[0,1]$ to obtain the ``working'' selection rule: $\xi_{\text{work}} = \max(0, \xi^*) / \max(\xi^*)$.
The order of the remaining units is determined by 
$\lambda \xi_{\text{work}} + (1-\lambda) \hat \delta$,  
where $\lambda \in [0,1]$ is the user-specified parameter that balances the power and diversity; we discuss the specific choices of $\lambda$ in our experiments in Section~\ref{sec:diversified_selection}.

It is straightforward to check that the ordering strategy is adapted to the filtration $\cF_\ell$, 
and therefore the resulting selection procedure still has its FDR controlled below $\alpha$ due to Theorem~\ref{thm:fdr_control}. 

\subsection{Incorporating new labels}
\label{sec:newlabels}

Our last instantiation concerns incorporating newly available labels from the screened samples. 
In the LLM deployment example, the test units that are ruled out by the selection process
may be sent to human annotators for labeling. 
Once available, these labels can be integrated into ACS
to improve the prediction model.

Formally, at step $\ell \ge k$, assume that we additionally 
have access to the label $Y_i$ for $i \in \cL_\ell \subseteq \{j \in O_\ell: A_j = 1\}$. 
We can then train the prediction model $\hat g_\ell$ via 
\$
\hat g_\ell = \cG(Z_{\pi(1)},\ldots,Z_{\pi(k)},\tZ_{\pi(k+1)},\ldots,\tZ_{\pi(\ell)}, 
X_{\pi(\ell+1)},\ldots,X_{\pi(n+m)}), 
\$
where $\tZ_i = Z_i$ if $A_i = 0$ or $i \in \cL_\ell$, and $\tZ_i = X_i$ otherwise. 
As before, the unit to be screened at step $\ell+1$ is chosen via
\$
\pi_{\ell+1} = \argmin{i \in U_\ell}~ \hat g_\ell(X_i).
\$
Although $\hat \pi_\ell$ is no longer adapted to the filtration $\cF_\ell$ as defined in \eqref{eq:filtration_alg}, it is adapted to a slightly richer filtration---which incorporates the new label information as it becomes available---for which the proof of Theorem~\ref{thm:fdr_control} still goes through. This is an implication of the 
proof of Theorem~\ref{thm:fdr_control}; see Appendix~\ref{appd:proof_fdr_control}.



\section{Numerical simulations}
\label{sec:simulation}
This section contains simulation studies to evaluate the performance of ACS in various settings. 
In Section~\ref{sec:sim_setup} we lay out the simulation setups.
Section~\ref{sec:base_learners} compares ACS and CS with various base learners,
and Section~\ref{sec:adaptive_model_selection} shows the performance of ACS with adaptive model selection;
Section~\ref{sec:newlabels_sim} and~\ref{sec:diversified_selection} demonstrate the performance of 
ACS for diversity-oriented selection and incorporating new labels, respectively.

\subsection{Simulation setup}
\label{sec:sim_setup}
Our simulation setups follow those in~\citet{jin2023selection}. 
The covariates $X_i$ are i.i.d.~sampled from $\text{Unif}[-1,1]^{20}$ 
and the response $Y_i$ is generated from the model: 
$Y_i = \mu(X_i)+\varepsilon_i$, where $\mu(x)$
is the true regression function, and $\varepsilon_i$'s are 
independent Gaussian noise terms.
Throughout this section, the true regression function is set to be
\$ 
\mu(x) = 4\ind\{x_1x_2>0\}\cdot\max\{x_4,0.5\} + 
4\ind\{x_1x_2 \le 0\}\cdot \min\{x_4,-0.5\}.
\$ 
The noise term $\varepsilon_i$  
is i.i.d.~from $\cN(0,\sigma^2)$, with 
$\sigma\in\{0.03,0.06,\ldots,0.15\}$ unless otherwise specified.
Appendix~\ref{appd:simulation} provides 
additional simulation results for other choices of $\mu(x)$ and the noise distribution.

We take the property set $\cC_{i} = (-\infty, 0]$, for $i\in[n+m]$. That is, 
the goal is to select units with $Y_{n+j} >0$. 
The number of test units is fixed to be $m=100$, except for the simulations in 
Section~\ref{sec:diversified_selection}, where $m=200$. 
The number of labeled units is $2n$ for both CS and~\algoname, with $n \in \{200,500,1000\}$.
For CS, $n$ units are used for training and the remaining $n$ for calibration;
in ACS, we similarly set $k = n$ to ensure a fair comparison. We consider gradient boosting (GB), random forest (RF), and support vector machine (SVM) 
as the model fitting function $\cG$, implemented via 
the \texttt{scikit-learn} Python package~\citep{pedregosa2011scikit}, 
without hyperparameter tuning. The target FDR level is set as $\alpha = 0.1$.

For each setting, we repeat the simulation $N = 1,\!000$ times, and report the average power, FDR, and ES. 
In specific, letting $\cS^{(i)}$ denote the selection set  in the $i^{\text{th}}$ simulation and 
$Y_{n+j}^{(i)}$ the $j^{\text{th}}$ test unit in the $i^{\text{th}}$ simulation, we compute the following quantities: 
\$ 
& \widehat{\text{Power}}  = \frac{1}{N}\sum_{i=1}^N \frac{\sum^m_{j=1} \ind\{Y^{(i)}_{n+j}>0, j \in \cS^{(i)}\}}
{\sum_{j=1}^m \ind\{Y^{(i)}_{n+j}>0\}}, \\ 
& \widehat{\text{FDR}}  = \frac{1}{N}\sum_{i=1}^N \frac{\sum^m_{j=1} \ind\{Y^{(i)}_{n+j}\le 0, j \in \cS^{(i)}\}}
{|\cS^{(i)}|}, \\
& \widehat{\text{ES}}  = \frac{1}{|\cN_2|}\sum_{i \in \cN_2} \frac{\sum_{1 \le j<k \le m} 
\ind\{j \in \cS^{(i)}, k \in \cS^{(i)}\} \cdot \ind\{Y_{n+j}^{(i)} >0, Y_{n+k}^{(i)} >0\} 
\cdot \theta(X^{(i)}_{n+j}, X^{(i)}_{n+k})}
{\sum_{1 \le j<k \le m} \ind\{j \in \cS^{(i)}, k \in \cS^{(i)}\} \cdot\ind\{Y^{(i)}_{n+j} >0, 
Y^{(i)}_{n+k} >0\}},
\$
where $\cN_2$ is the set of simulations with at least two units selected.

\subsection{ACS with model-refitting: individual base models}
\label{sec:base_learners}
We first compare ACS with CS under different fixed base models $\cG$, following the procedure in the ``Model re-fitting'' part of Section~\ref{sec:power}. 
In this case, we re-fit the prediction model---within a fixed model class---every $L=10$ steps with all labeled samples that have been screened up to step $\ell$; 
the ordering of the unscreened samples is then determined by the most recently re-fitted model and remains fixed until the next model update.

\begin{figure}[htbp!]
\centering
\includegraphics[width=\textwidth]{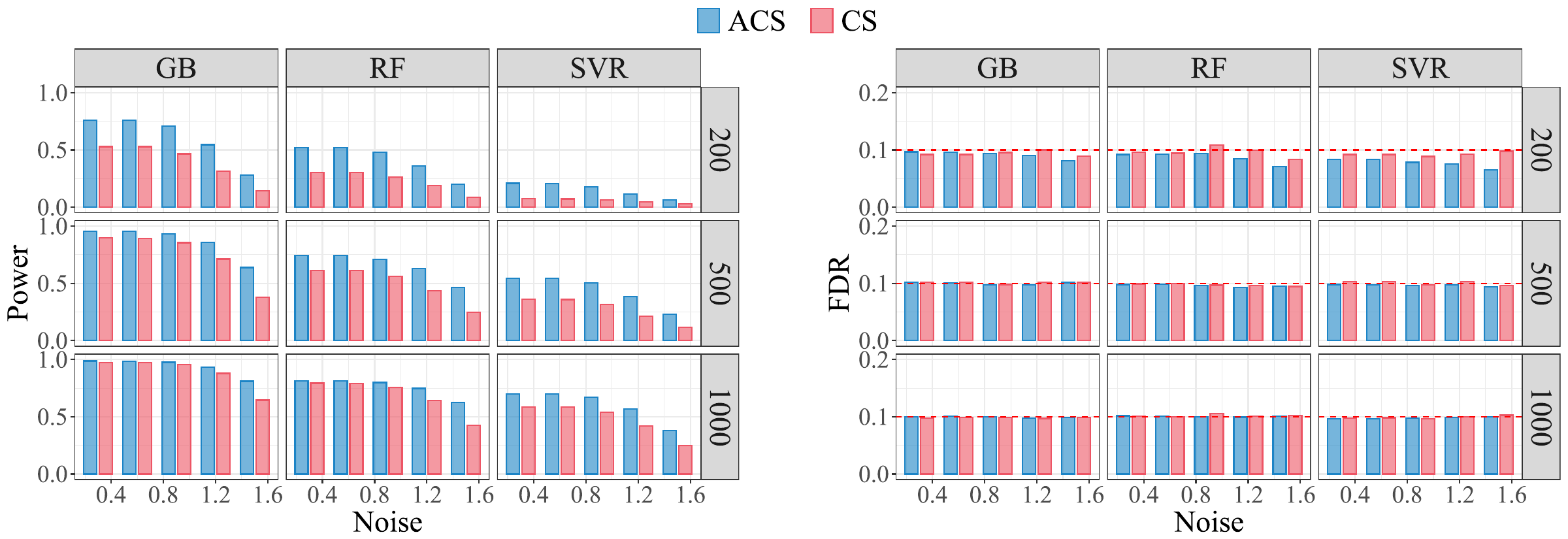} 
\caption{Realized power (left) and FDR (right) of ACS and CS 
for various noise levels $\sigma$. 
Each column corresponds to one of three fixed base models: gradient boosting (GB), random forest (RF), 
and support vector machine (SVR). Each row corresponds to one sample size $n\in \{200,500,1000\}$. The target FDR level is $\alpha = 0.1$ 
and the results are averaged over $1,\!000$ independent runs.} 
\label{fig:set1_simulation}
\end{figure}

We fix $\cG \in \{\text{GB}, \text{RF}, \text{SVR}\}$
and $n \in \{200,500,1000\}$. Figure~\ref{fig:set1_simulation} shows the realized power and FDR of CS and ACS.
Both methods achieve the target FDR level $\alpha = 0.1$ in all settings, confirming the theoretical results. 
For any choice of $\cG$ and any $n$, ACS achieves higher power than CS, and the improvement is more significant 
when $n$ is smaller and/or the signal-to-noise ratio is lower.
The power gain of ACS over CS in such cases is primarily due to the increased number of labeled data for  
model fitting, which leads to a more accurate prediction model and hence a better ordering
of the units.

\subsection{ACS with model-refitting: adaptive model selection}
\label{sec:adaptive_model_selection}    
Under the same setup of Section~\ref{sec:base_learners}, we evaluate the performance of 
ACS using the procedure described in the ``Adaptive model selection'' part of Section~\ref{sec:power}. 
In our implementation, at every $L=20$ steps, ACS adaptively chooses from the candidate 
models $\{\text{GB}, \text{RF}, \text{SVR}\}$ as the working model, and re-fit 
the chosen model with all the labeled data screened up to the currents step.
The ordering of the unscreened samples is then determined 
by the most recently updated model and remains fixed until the next update. 
To study the impact of incorporating additional labeled samples into the filtration, 
we also implement a heuristic approach that selects the model 
based on the estimated mean squared error (MSE) of the fitted model computed 
on the training data---that is, both model fitting and MSE computation are performed
using $\cD_\tr$.
We refer to this 
approach as the ``naive'' method, which allows for model selection (on the training data) but does not leverage additional labeled data.

\begin{figure}[htbp!] 
\centering
\includegraphics[width=0.9\textwidth]{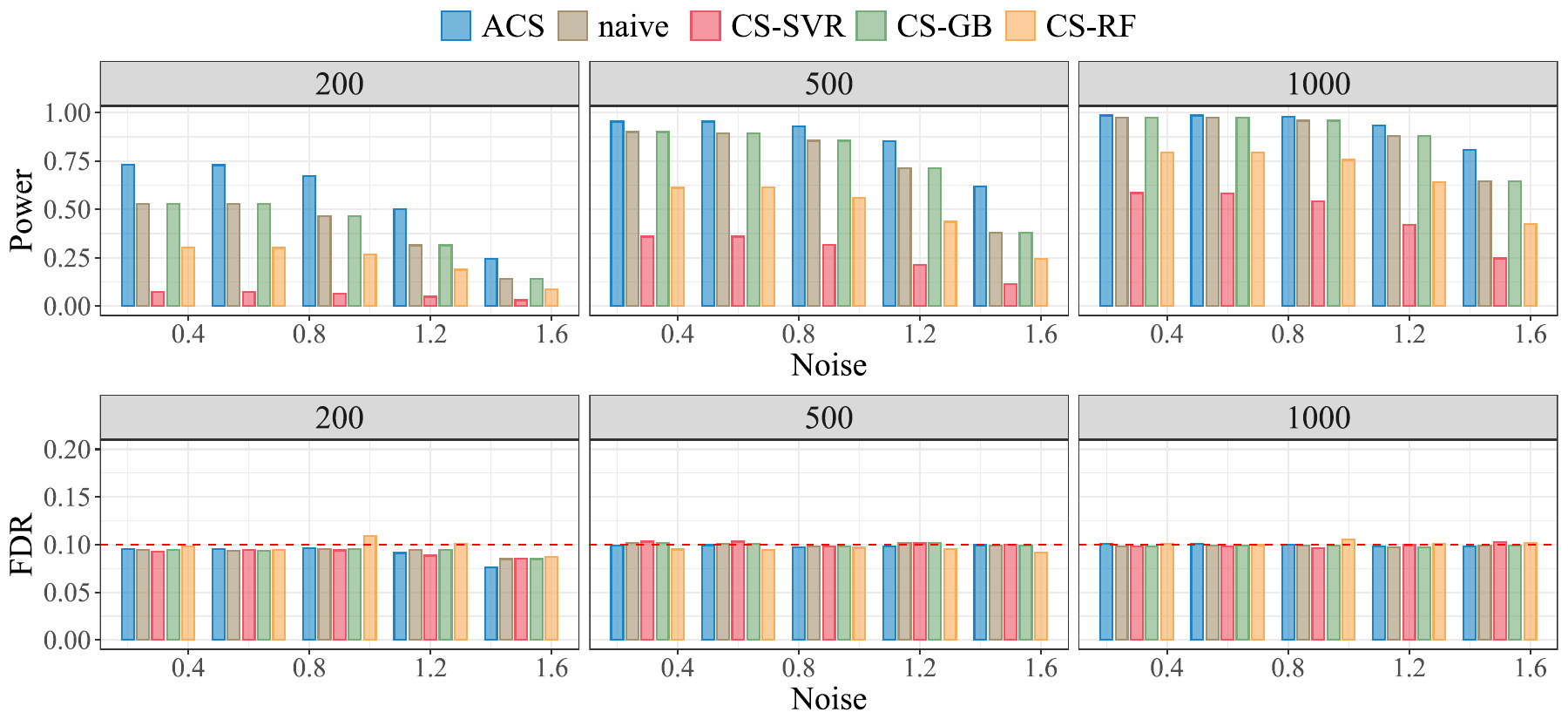} 
\caption{Realized power (top) and FDR (bottom) of ACS with adaptive model selection, 
the naive method, CS-SVR, CS-RF, and CS-GB, as a function of the noise level $\sigma$. 
The target FDR level is $\alpha = 0.1$ 
and the results are averaged over $1,\!000$ independent simulations.}
\label{fig:ensemble_set1_simulation}
\end{figure}

Figure~\ref{fig:ensemble_set1_simulation} plots the realized power and FDR curves 
of ACS with adaptive model selection, the naive method, CS-SVR, CS-GB, and CS-RF.
It can be observed that ACS with adaptive model selection outperforms all the CS-based 
methods---as well as the naive approach---in terms of power, while maintaining the target FDR level.

As in the previous case, the power gain of ACS over CS is more significant when 
$n$ is smaller and/or the signal-to-noise ratio is lower.
With adaptive model selection, ACS can effectively leverage the strengths of different 
base models and adapt to the data structure without compromising the FDR control. 
The improvement in power is a combination of choosing the best model for the data 
and the increased number of labeled data for model fitting.

\subsection{Diversified selection}
\label{sec:diversified_selection}

We now study the performance of 
ACS with diversified selection, as introduced in Section~\ref{sec:diversified_selection}.
In this experiment, we take the number of test samples to be $m=200$ and 
$n \in \{500,1000\}$; the base model is set to be gradient boosting (GB), and the noise 
level ranges in $\{0.3,0.6,\ldots,3\}$. 
For the diversity measure $\theta$, we adopt the
the RBF kernel with $\sigma_0 = 5$. The 
other setups are the same as in the previous sections. 

For ACS, we implement the diversity-aware version, where
the ordering at each step $\ell \ge k$ is decided by the ``mixture'' score proposed in Section~\ref{sec:diversified_selection}: 
\$
\pi_{\ell+1} = \argmin{j \in U_{\ell}}~\lambda \xi_{\text{work}_j} + (1-\lambda) \hat \delta_j,
\$ 
where $\lambda \in [0,1]$ is the user-defined parameter that controls  
the relative weight of diversity over power. In this experiment, we set $\lambda = 0.3$; 
results for other choices of $\lambda$ can be found in Appendix~\ref{appd:simulation}, 
and panel (b) of Figure~\ref{fig:ensemble_roc} corresponds to $\lambda=1$.
The score is updated every $L=10$ steps.
We compare the performance of diversity-aware ACS with the power-oriented ACS (with $\lambda = 0$) and CS. 

\begin{figure}[h!]
\centering
\includegraphics[width=.6\textwidth]{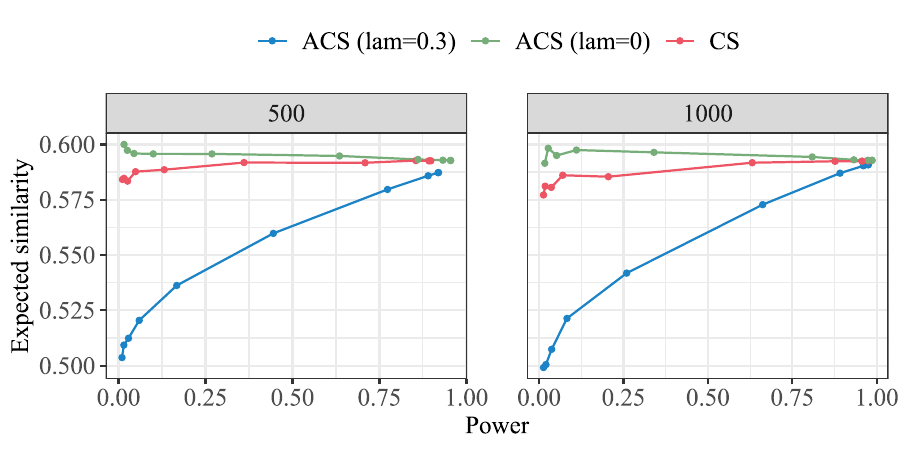}
\caption{$\widehat{\text{ES}}$ as a function of $\widehat{\text{Power}}$ for 
diversity-aware ACS with $\lambda = 0.3$, ACS, and CS. 
The noise level $\sigma$ is varied in $\{0.3,0.6,\ldots, 3\}$ and 
the results are averaged over $1,\!000$ independent simulations.}
\label{fig:div_set_es_1_lam_0.3_curve}
\end{figure}

\begin{figure}[h!]
\centering
\includegraphics[width=0.8\textwidth]{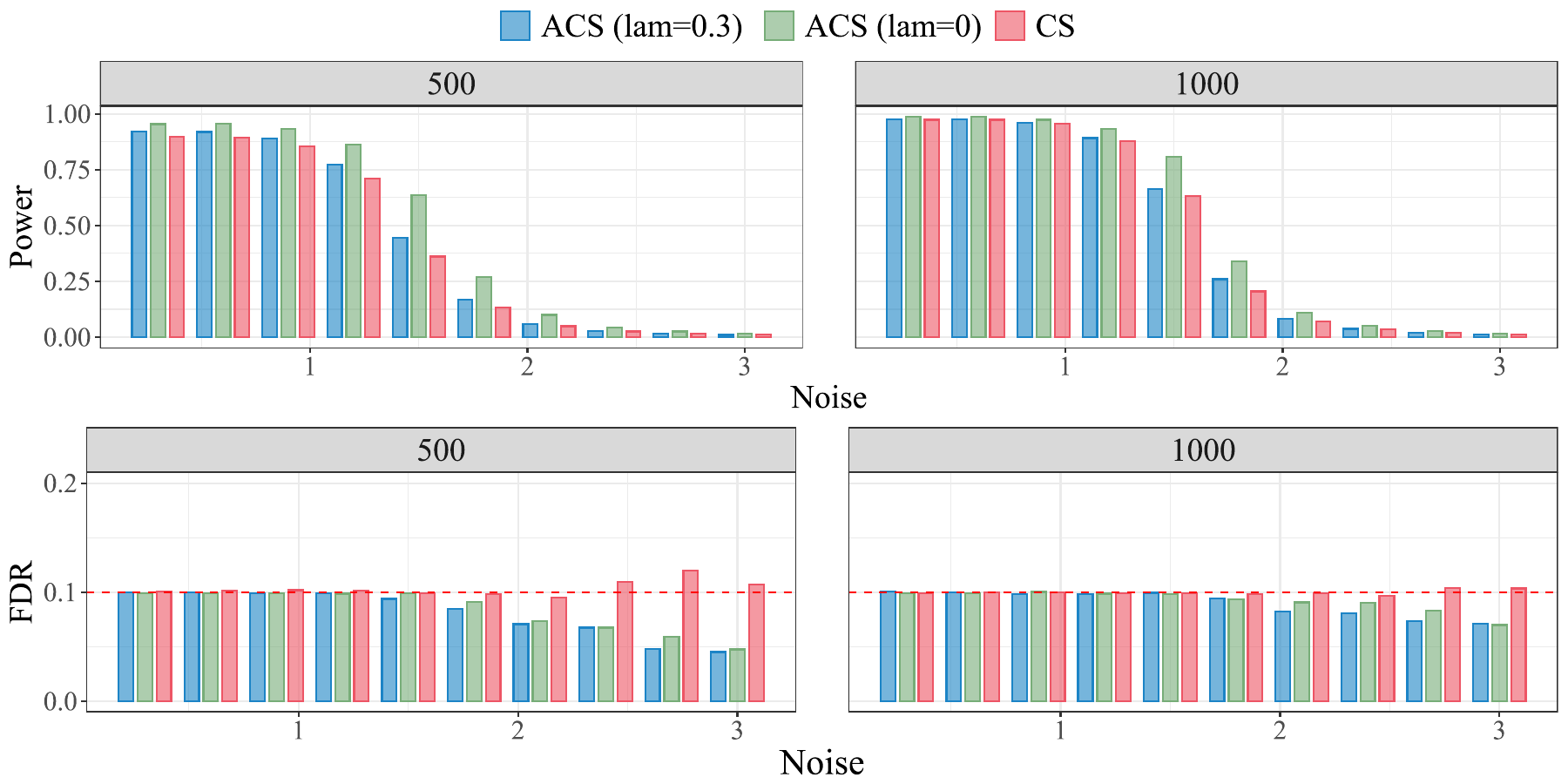}
\caption{Realized power (left) and FDR (right) as a function of the noise level $\sigma$ for 
diversity-aware ACS with $\lambda = 0.3$, ACS, and CS. 
The other details are the same as in Figure~\ref{fig:div_set_es_1_lam_0.3_curve}.}
\label{fig:div_set_es_1_lam_0.3}
\end{figure}

Figure~\ref{fig:div_set_es_1_lam_0.3_curve} plots the ES-Power curves for the three methods. 
The realized ES of diversity-aware ACS is substantially lower than the other two methods at each power level,  
demonstrating the overall improvement in diversity. Figure~\ref{fig:div_set_es_1_lam_0.3}
shows the realized power and FDR of the three methods at 
various noise levels. As expected, the FDR of all methods is controlled at the target level $\alpha = 0.1$; 
the power of diversity-aware ACS is a bit lower than that of ACS, which is also reasonable 
since it refrains from making a selection that is similar to the other selected. Perhaps a bit surprisingly though, 
diversity-aware ACS achieves a higher power than CS, while promising a more diversified selection. This shows that---through iteratively re-fitting the model and updating the scoring function---ACS finds better orderings of the unscreened samples that not only improve the power but also better discern the null and non-null samples.

\subsection{Incorporating new labels}
\label{sec:newlabels_sim}

\begin{figure}[htbp!]
\centering
\includegraphics[width=\textwidth]{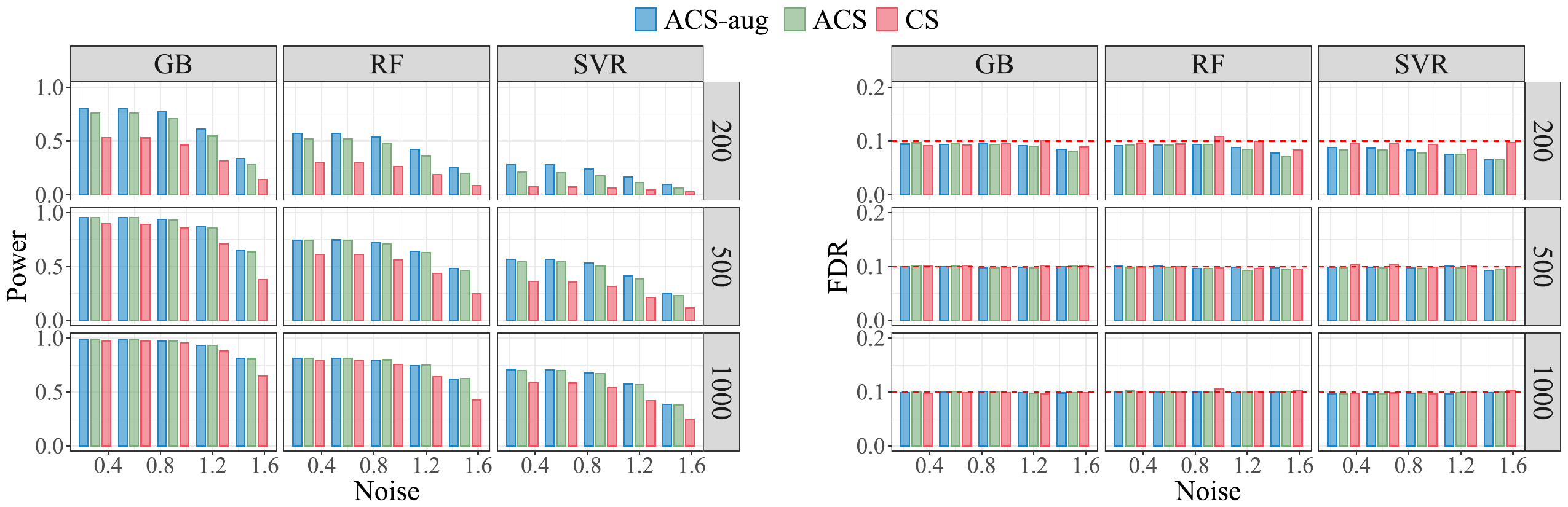}
\caption{Power (left) and FDR (right) of ACS with new labels (ACS-aug), 
ACS without new labels (ACS), and CS as a function of the noise level $\sigma$.
The simulation results correspond to the setting in Section~\ref{sec:newlabels_sim}, 
where the true labels of the test units are revealed after being screened.
The target FDR level is $\alpha = 0.1$ and the results are averaged over $1,\!000$ independent simulations.}
\label{fig:aug_set1_simulation}
\end{figure}

We now study the procedure in Section~\ref{sec:newlabels}, in a setting where the true labels of the test units become available {\em after} they 
are screened (i.e., knowing that they are not selected). Here, ACS leverages 
the new labels in 
re-fitting the model for every $L=10$ steps---that is, 
the model is fitted with $\{(Z_i,A_i)\}_{i\in O_\ell}$.
Figure~\ref{fig:aug_set1_simulation} shows the realized power and FDR of ACS that 
incorporates the new labels (referred to as ACS-aug), as well as those of the original ACS and CS. 

ACS-aug further improves the power of ACS, as more labeled data are available for model fitting. 
The FDR control is preserved in all settings, and the improvement in power is, as before, more significant 
when $n$ is smaller and/or the signal-to-noise ratio is lower.

\section{Real data application: LLM deployment}
\label{sec:real_data}

In this section, we demonstrate the application of ACS to 
selective deployment of foundation model outputs, following the setup of~\citet{gui2024conformal}. 
Here, we view a prompt-output 
pair as a unit, and seek to select a subset of prompts whose outputs are aligned with human evaluations.
Two specific tasks are considered: question answering and chest X-ray (CXR) report generation.

\paragraph{Background.}
In both examples, we let $E_i$ denote the expert (gold-standard) input corresponding to a prompt $X_i$.
For example, if $X_i$ is a question, $E_i$ is the true answer, 
and if $X_i$ is a medical image, $E_i$ is the radiology report written by a medical expert. The $E_i$'s are only available in the labeled dataset.

Given a foundation model $f\colon \cX\to \cY$ that generates an output $f(X)\in \cY$ given an input prompt $X\in \cX$, we assume access to an evaluation function $\mathcal{A}: \cY \times \cY \mapsto \{0,1\}$ 
that takes the foundation model output 
$f(X_i)$ and the expert input $E_i$ as inputs, and outputs a binary alignment score 
$Y_i = \mathcal{A}(f(X_i), E_i) \in \{0,1\}$. The specific form of the evaluation function
$\mathcal{A}$ will be specified in each task. 
The alignment score $Y_i=1$ indicates that the output $f(X_i)$ is aligned with the expert input $E_i$, 
while $Y_i=0$ indicates misalignment.
Since the $E_i$'s are only available in $\cD_\lab$, we 
only observe the alignment scores $Y_i$ in $\cD_\lab$; the goal 
is to select a subset of the test units with $Y_{n+j}=1$.
In other words, the null hypothesis of interest takes the form $\mathcal{H}_{0,j}: Y_j \in \cC_j$ with $\cC_j = \{0\}$.

\paragraph{Experimental setup.}
We fix the size of the test set at $m=|\cD_\te|=1000$ 
and the set of labeled samples (those with expert output) is of size $N$; 
the latter is further partitioned into three subsets $\cD_\tr \cup \cD_{\rm tune} \cup \cD_{\cal}$. 
Throughout, we fix $|\cD_{\tr}|/N = 0.1$, $|\cD_{\text{tune}}|/N = 0.2$, 
and vary $N$ in $\{100,500,1000\}$. 

Following~\citet{gui2024conformal}, to build the prediction model, we use $\cD_{\rm tune}$ for feature engineering, where we 
extract certain features from the foundation model outputs---such as self-evaluation likelihoods~\citep{kadavath2022language,lin2023generating}, input uncertainty scores~\citep{kuhn2023semantic, lin2023generating}, and output confidence scores~\citep{lin2023generating,gui2024conformal}---as the covariates $\tilde X_i$. 
Based on the extracted covariates $\tilde X_i$, we fit the model for predicting the alignment scores $Y_i$ using $\cD_{\tr}$, 
with base classifiers including logistic regression (\texttt{logistic}), random forests (\texttt{RF}), and XGBoost (\texttt{XGBRF}). 
More details are deferred to Appendix~\ref{sec:additional_llm}. 

\subsection{Experiments for question answering (QA) datasets}\label{sec:exp_qa}
We begin with the question answering task, where we select question answers that are sufficiently close to the correct answers in two widely adopted datasets: a conversational question answering dataset {\bf TriviaQA} \citep{joshi2017triviaqa} and a closed-book reading comprehension dataset {\bf CoQA} \citep{reddy2019coqa}. 
To investigate the performance across LLMs with different capabilities, 
we employ two LLMs (without finetuning)   {\bf OPT-13B}~\citep{zhang2023opt} and 
{\bf LLaMA-2-13B-chat}~\citep{touvron2023llama} to generate an answer $f(X_i)$ given the input questions $X_i$.\footnote{Generations are based on top-p sampling for each input $X_i$ following the default configuration.} The alignment score $Y_i$ is chosen to be $\mathbf{1}\{\texttt{rouge-L}_i \geq 0.3\}$ following \cite{kuhn2023semantic,lin2023generating,gui2024conformal}, where the \texttt{rouge-L} score \citep{lin2004rouge} measures the similarity between generated and correct answers based on the longest common subsequences. 

\begin{figure}[h]
\centering
\includegraphics[width=\textwidth]{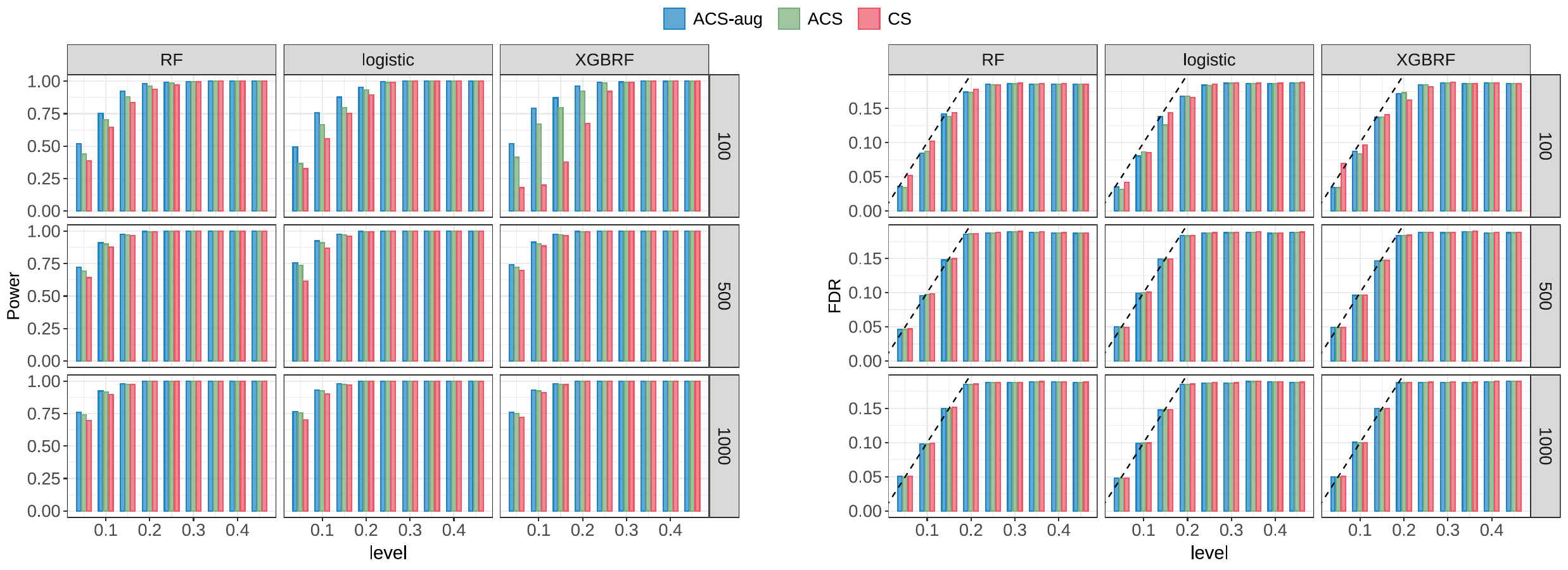} 
\caption{Realized power (left) and FDR (right) of ACS with new labels (ACS-aug), ACS without new labels
(ACS), and CS from experiments on the CoQA dataset with LLaMA-2-13B-chat as 
the base LLM. Each column corresponds to a model class and each row corresponds to the size $N$ of labeled samples. 
The target FDR level is $\alpha = 0.1$ 
and the results are averaged over $200$ independent simulations.}
\label{fig:aug-lm-coqa}
\end{figure}

\paragraph{ACS and ACS-aug with individual base models.}
Following the setup in Section~\ref{sec:base_learners}, 
we implement both CS and ACS, fixing $\cG \in \{\texttt{logistic}, \texttt{RF}, \texttt{XGBRF}\}$. 
Figure~\ref{fig:aug-lm-coqa} presents results on the CoQA dataset 
with LLaMA-2-13B-chat as the base language model;
results for OPT-13B and the TriviaQA dataset are deferred to Appendix~\ref{sec:additional_llm}.
As shown in the figures, both ACS and CS successfully 
control the FDR across all base classifiers. 
However, ACS consistently achieves substantially higher power, 
particularly at lower nominal FDR levels. 
The power advantage of ACS is especially pronounced 
when $N=100$ and \texttt{XGBRF} is the base classifier, 
highlighting the power advantage of ACS in settings with 
limited samples and data-consuming predictive models. 

We further consider the scenario in Section~\ref{sec:newlabels_sim} where alignment scores are revealed after screening, in which case, the effective sample size keeps increasing and can be used for updating the prediction model in the ACS framework;
we refer to this variant as ACS-aug. In our experiment, we fix the model class 
for ACS-aug, and retrain the model using all labeled, screened data every 
$L=20$ steps. The results for ACS-aug, also shown in Figure~\ref{fig:aug-lm-coqa}, demonstrate further improvements in power---particularly in settings with a limited number of labeled samples (i.e., when  $N$ is small). 

\begin{figure}[h]
\centering
\includegraphics[width=\textwidth]{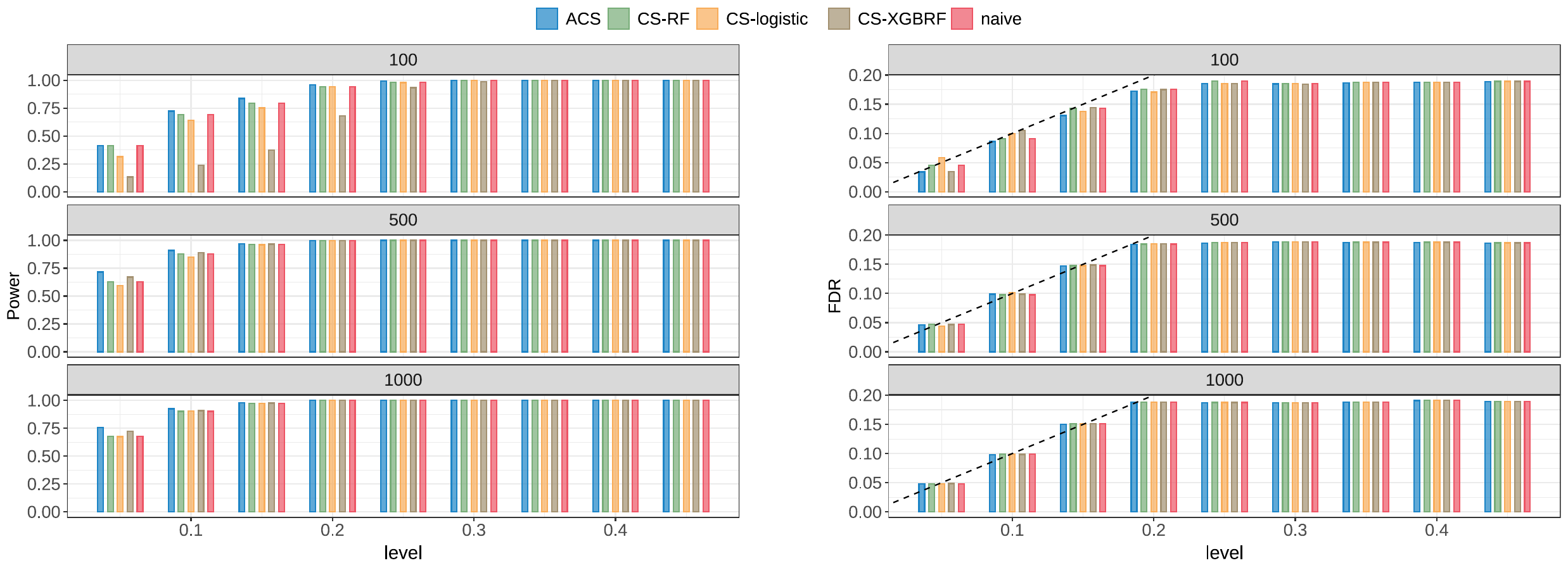} 
\caption{Realized power (left) and FDR (right) of ACS with adaptive model selection,  the naive method, CS-SVR, CS-RF, and CS-GB  from experiments on the CoQA dataset with LLaMA-2-13B-chat as the base LLM.
Each subplot corresponds to the size $N$ of labeled samples.
The target FDR level is $\alpha = 0.1$ 
and the results are averaged over $200$ independent simulations.}
\label{fig:en-lm-coqa}
\end{figure}

\paragraph{ACS with adaptive model selection.}
Following the pipeline in Section~\ref{sec:adaptive_model_selection}, 
we evaluate ACS with adaptive model selection, the naive method, CS-logistic, CS-RF, and CS-XGBRF.
In this experiment, ACS adaptively selects the prediction model $\cG \in \{\texttt{logistic}, \texttt{RF}, \texttt{XGBRF}\}$ every $L=20$ steps.
We plot the realized power and FDR in Figure~\ref{fig:en-lm-coqa}
from experiments on the CoQA dataset with LLaMA-2-13B-chat as the base LLM
(results for TriviaQA are deferred to Appendix~\ref{sec:additional_llm}). 
In both cases, we see that ACS with adaptive model selection
achieves the highest power compared to the baselines while controlling FDR at the target level. 
Again, the advantage is more pronounced when the sample size $N$ is small, 
or when the target FDR level is stringent. 


\subsection{Experiments for chest X-ray report generation}
We now turn to the task of chest X-ray report generation, 
where we aim to select a subset of machine-generated radiology reports 
that are aligned with expert evaluations.
We evaluate the performance of ACS on the chest X-ray (CXR) dataset, 
which is a subset of the MIMIC-CXR dataset~\citep{johnson2019mimic} consisting of chest radiographs 
and radiology reports. 
Given an input image $X_i$, a fine-tuned vision language model (VLM) 
with a pre-trained VisionTransformer as image encoder and the GPT2 as the text generator 
is used to generate radiology reports, $f(X_i)$. 
To measure the alignment between VLM-generated and expert-written reports, 
we use CheXbert~\citep{smit2004chexbert} to map both kinds of reports to $14$-dimensional categorical vectors and the alignment score $Y_j=1$ if and only if there are at least $12$ matched labels. Base classifers and experimental setups are the same with Section~\ref{sec:exp_qa}.

\paragraph{ACS and ACS-aug with individual base models.}

\begin{figure}[h]
\centering
\includegraphics[width=\textwidth]{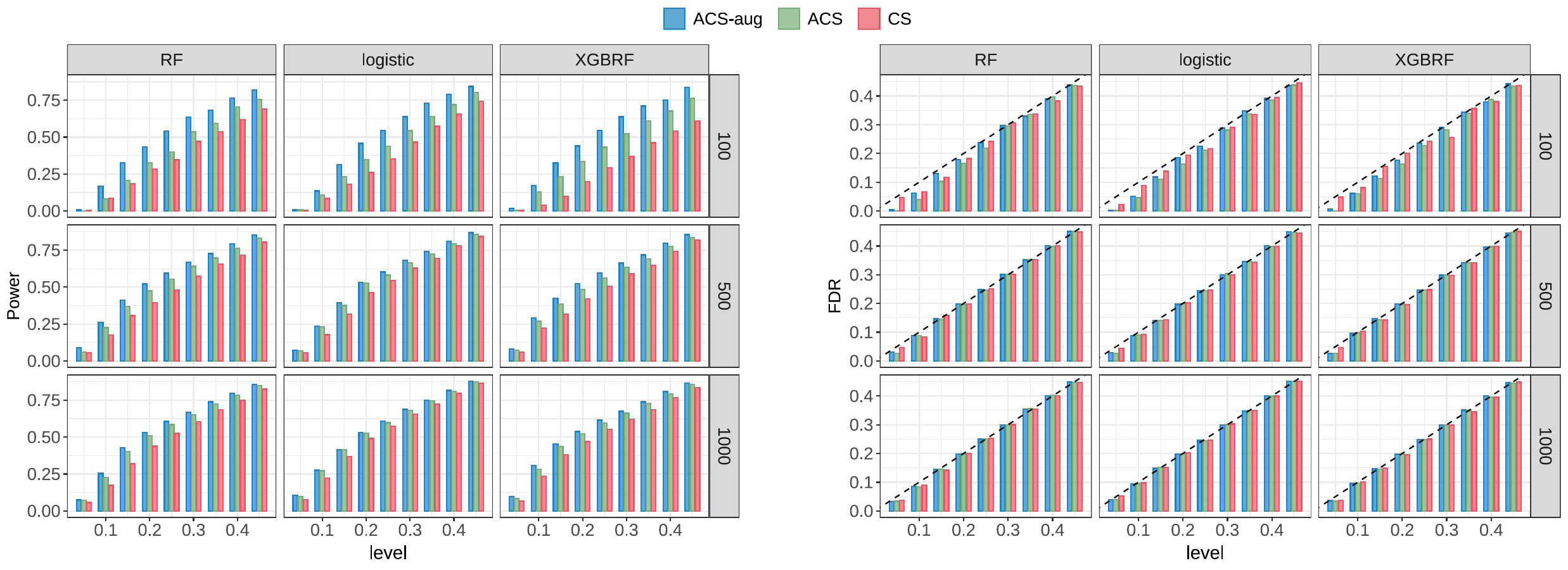} 
\caption{Realized power (left) and FDR (right) of ACS with new labels (ACS-aug), ACS without new labels
(ACS), and CS from experiments on the MIMIC-CXR dataset. 
Each column corresponds to a model class and each row corresponds to the size $N$ of labeled samples.
The target FDR level is $\alpha = 0.1$ 
and the results are averaged over $200$ independent simulations.}
\label{fig:aug-cxr}
\end{figure}

In Figure~\ref{fig:aug-cxr}, we compare the performance of CS, ACS, and ACS-aug across different base classifiers: \texttt{RF}, \texttt{logistic}, and \texttt{XGBRF}. In this limited-data setting, ACS consistently achieves higher power than CS, and ACS-aug further boosts power by increasing the effective sample size. The performance gain is particularly obvious when the sample size $N$ is small and the classifier is more data-intensive, such as \texttt{XGBRF}.

\begin{figure}[h]
\centering
\includegraphics[width=\textwidth]{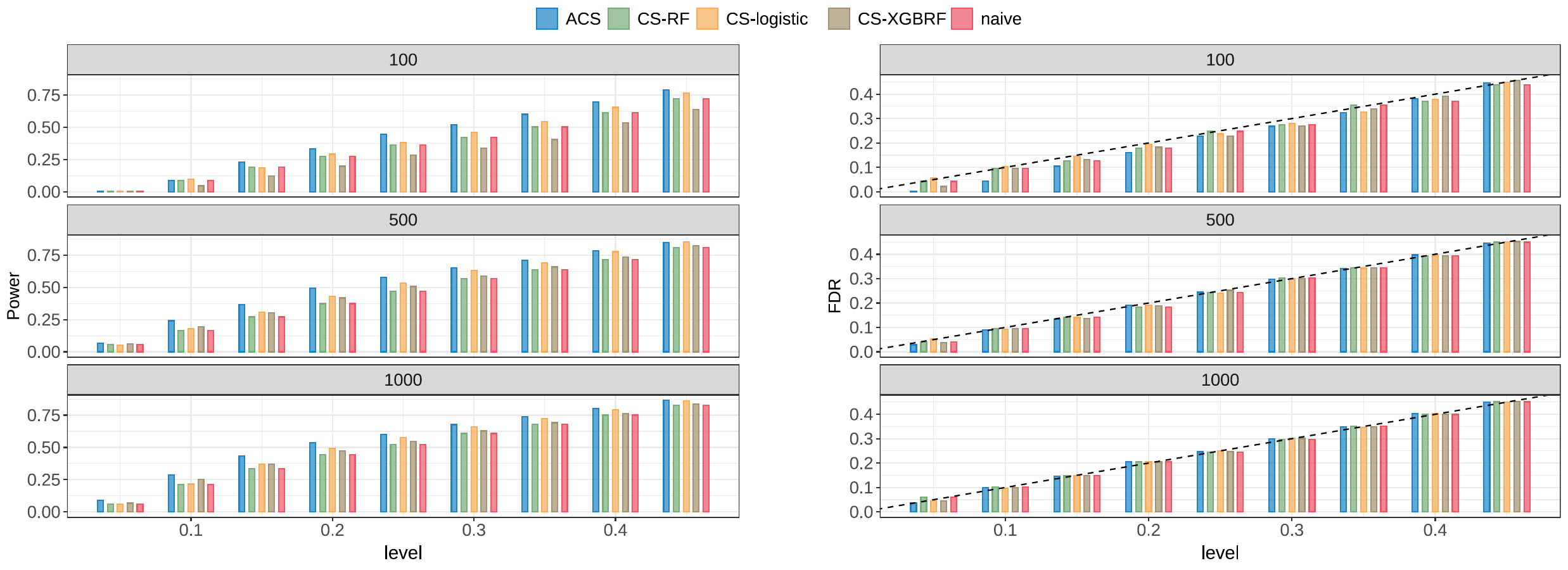} 
\caption{Realized power (left) and FDR (right) of ACS with adaptive model selection, 
the naive method, CS-SVR, CS-RF, and CS-GB: MIMIC-CXR dataset. 
Each subplot corresponds to the size $N$ of labeled samples.
The target FDR level is $\alpha = 0.1$ 
and the results are averaged over $200$ independent simulations.}
\label{fig:en-cxr}
\end{figure}

\paragraph{ACS with adaptive model selection.}
We further enhance the performance of ACS via adaptive model selection following Section~\ref{sec:adaptive_model_selection} and compare its performance with CS using fixed base classifiers in Figure~\ref{fig:en-cxr}.
In this task, the performance of the base models are largely similar, so the naive approach leads to little improvement. However, ACS with adaptive model selection still achieves visible power improvement while tightly controlling the FDR. 

\section{Real data application: Drug discovery}
\label{sec:drug_discovery}

In this section, we use ACS to make diverse selections on a drug discovery dataset, 
where we will follow the notation in Section~\ref{sec:diversity-aware-ordering}. 
Our experiments are performed on a catalytic receptor interaction dataset from \cite{drug_discovery}, 
consisting of ligand-receptor pairs ($X$) along with the binding affinity ($Y$) of each pair; large binding affinities indicate molecular potency, 
and the goal is to select a subset of ligand-receptor pairs  with high binding affinity. 
Our goal will be to return a selection that is diverse as measured by the chemical similarity between ligands. More specifically, writing the pair as $X=(X_{\mathrm{ligand}}, X_{\mathrm{receptor}})$, we aim to return a selection set that contains a diverse set of $X_{\mathrm{ligand}}$'s. To this end, we measure similarity between the $X_{\mathrm{ligand}}$'s via the Tanimoto similarity \citep{tanimoto1958elementary,bajusz2015tanimoto} between their fingerprints.
Following~\cite{jin2023selection}, we consider the property sets $\mathcal{C}_i=(-\infty, c_i]$ 
where $c_i$ is the $q$-quantile of binding affinities in the training set with the same target receptor as that of the $i^{\text{th}}$ example.

The machine learning algorithm we use to make quality predictions is a small neural network architecture. This model is a regressor and as such, we obtain the property set non-inclusion predicted probabilities $\hat \delta_j \in [0,1]$ by taking softmax. More specifically, if the remaining candidates are $i=1, \ldots, \tilde{n}$, 
then we take \[\hat \delta_i = \frac{\exp(\hat{g}(X_i))}{\sum_{j=1}^{\tilde{n}}\exp(\hat{g}(X_j))}, \quad i=1, \ldots, \tilde n,\] where $\hat{g}$ is the fitted neural network regressor at the current screening step. 
 
To proceed, we use a variant of the diversity-aware screening introduced in Section~\ref{sec:diversity-aware-ordering} with the FDR constraint turned off:
\begin{equation}
\begin{aligned}
\label{eq:diversity_opt_sample_drug}
\min_{\xi }~ &\frac{\sum_{i,j=1}^{\tilde n}
\xi_i \xi_j \hat\delta_i \hat \delta_j \cdot \theta(X_i,X_j)}
{\sum_{i,j=1}^{\tilde n}\xi_i \xi_j \hat \delta_i \hat \delta_j}\\
\text{s.t.}~& 0 \leq \xi_i \leq 1, \quad i=1, \ldots, \tilde{n}
\end{aligned}
\end{equation}
Using a similar proof to that of Proposition~\ref{prop:diversity}, it can be shown that the program~\eqref{eq:diversity_opt_sample_drug} can be reduced to 
\begin{equation}
\begin{aligned}
\label{eq:diversity_opt_sample_drug_2}
\min_{\xi }~ &\xi^\top  \Theta \xi\\
\text{s.t.}~& \xi_i \ge 0, \quad i=1, \ldots, \tilde{n}\\
&\xi^\top\hat \delta = 1
\end{aligned}
\end{equation} which is a quadratic program and can be efficiently solved.
Finally, just as in Section~\ref{sec:diversified_selection}, we actually determine the order of the remaining units by $\lambda \xi_{\mathrm{work}} + (1-\lambda)\hat{\delta}$ and take $\lambda=0.8$ in our experiments.

Our results are averaged over $250$ random train/calibration/test splits of sizes $113$, $113$, and $225$, respectively and the version of CS to which we compare uses the p-value described in Section~\ref{subsec:cs} (as opposed to those involving uniform tie-breaking random variables). 

\begin{figure}[h]
\centering
\includegraphics[width=\textwidth]{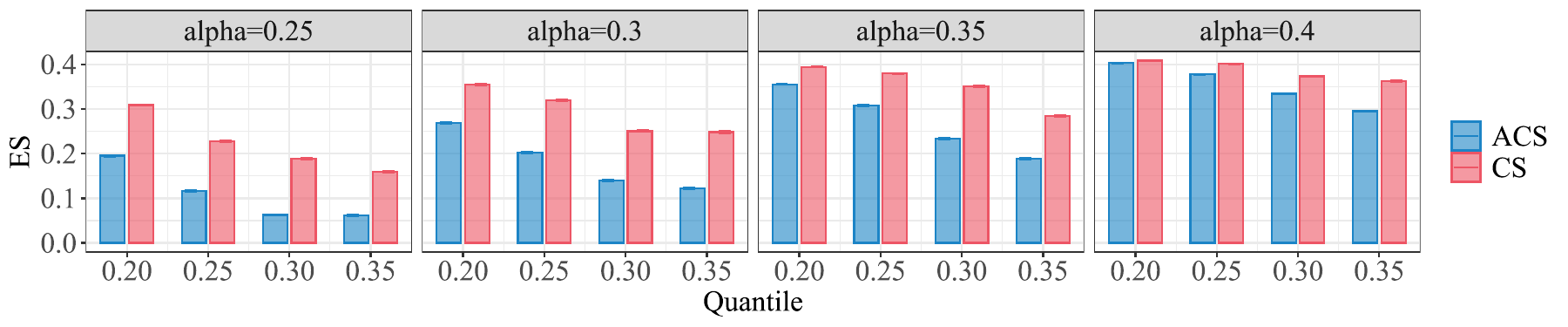} 
\caption{ES of ACS (blue) compared to CS (red) for various quantile values $q$ and nominal FDR levels $\alpha$ on real drug discovery dataset. 
The results are averaged over $250$ random train/calibration/test splits.}
\label{fig:drug-ess}
\end{figure}

\begin{figure}[h]
\centering
\includegraphics[width=\textwidth]{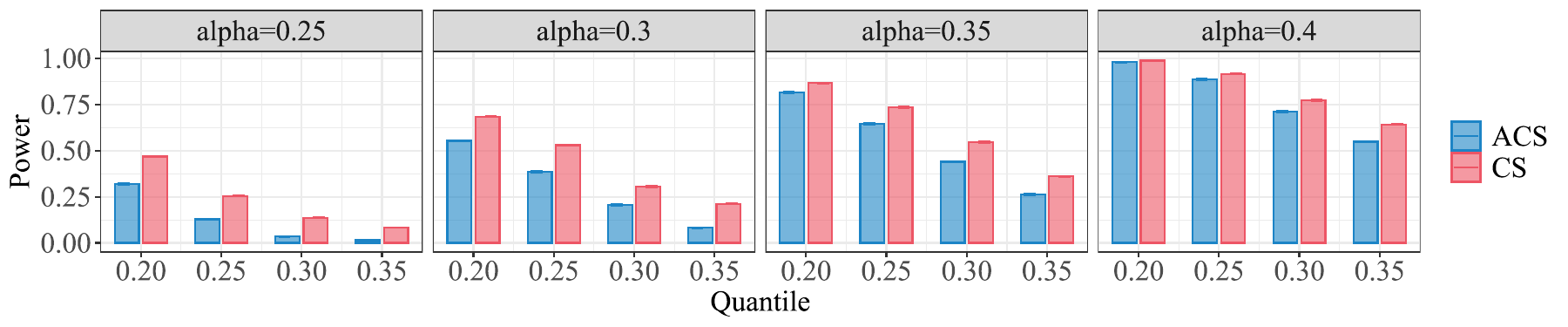} 
\caption{Power of ACS (blue) compared to CS (red) for various quantile values $q$ and nominal FDR levels $\alpha$ on real drug discovery dataset. 
The other details are the same as in Figure~\ref{fig:drug-ess}.}
\label{fig:drug-power}
\end{figure}

Figure~\ref{fig:drug-ess} shows the ES of our method compared to CS for various values of $q \in \{0.2, 0.25, 0.3, 0.35\}$ 
as well as the nominal FDR level $\alpha \in \{0.25, 0.3, 0.35, 0.4\}$. 
The general takeaway is that ACS with diversity-aware ordering produces selection sets with lower expected similarity than CS. 
Interestingly, the gap between the effective similarities of ACS and CS shrinks as both the nominal level $\alpha$ increases and the quantile $q$ decreases. 
This is most likely a reflection of the fact that the selection problem becomes easier for large $\alpha$ and small $q$, 
thereby allowing the FDP estimator to more easily fall below $\alpha$ at early timesteps. 
Indeed, this is reflected by the power plot given in Figure~\ref{fig:drug-power}, 
which shows that the power of both methods in these settings is either equal to or very near one. 
We defer results showing the FDR of both methods to Appendix~\ref{appendix:drug-discovery}.

\section{Discussion}
\label{sec:discussion}
This work introduces ACS, a model-free selection algorithm offering 
both flexibility and rigorous control of FDR.
The proposed framework allows for dynamic model update and selection, 
incorporating new information to maximize the selection power, 
and/or adapting to other utility functions determined by the practitioners. 
We demonstrate the performance of ACS in a variety of simulated settings and 
two real applications, showing that it outperforms existing methods 
in multiple dimensions. We conclude this paper by discussing some interesting 
directions for future research. 

\paragraph{Extension to outlier detection.}
The idea behind the proposed ACS framework can be extended to the outlier detection problem, 
where the goal is to identify a set of outliers from a given dataset, relative 
to a set of inliers~\citep{bates2023testing,marandon2024adaptive}. In outlier detection problems,  
it would be interesting to investigate the modelling approaches that  are beneficial for 
the selection of outliers.

\paragraph{Other utility functions.}
This work mainly focuses on maximizing the power and/or  diversity 
of the selected set. Given the flexibity of our framework, it is 
possible to aim for other utilities that appear in different applications
(e.g., achieving conditional guarantees, minimizing user-specified risk functions). 

\paragraph{Fully online selection.}
One limitation of the current ACS framework is that it does 
not accommodate the scenario where the test data arrives sequentially.
An interesting direction for future research is to study an online 
version of ACS that can adaptively update the model and incorporate new information
while maintaining appropriate type-I error controls.

\subsection*{Reproducibility}
Code for reproducing the numerical results in this paper is available at \url{https://github.com/zhimeir/acs_paper}.

\subsection*{Acknowledgments}
The authors would like to thank the Wharton Research Computing team for the computational resources
provided and the great support from the staff members.  
Z.~R.~is supported by the National Science Foundation (NSF) under grant DMS-2413135
and Wharton Analytics. Y.~N.~acknowledges support by a Graduate Research Fellowship from the National Science Foundation.
\newpage
\bibliographystyle{apalike}
\bibliography{ref}

\newpage
\appendix
\section{Proofs of the main results}
\subsection{Proof of Theorem~\ref{thm:fdr_control}}
\label{appd:proof_fdr_control}
For notational convenience, we further define 
$P_\ell^- = P_\ell \cap \cH_0$ and $P_\ell^+ = P_\ell \backslash P_\ell^-$.
By the definition of FDR and the choice of the stopping time $T$, we have
\$
\fdr  = \EE\Bigg[\frac{|P^-_{T}|}{|P_T| \vee 1}\Bigg]
&= \EE\Bigg[\frac{m}{n-k+1}\frac{1 + |N^-_T|}{|P_T| \vee 1} 
\cdot  \frac{n-k+1}{m}\frac{|P^-_T|}{1+|N^-_T|}\Bigg]\\
&= \EE\Bigg[ \widehat{\text{FDP}}(T) 
\cdot  \frac{n-k+1}{m}\frac{|P^-_T|}{1+|N^-_T|}\Bigg]\\
& \le \frac{\alpha(n-k+1)}{m} \EE\Bigg[\frac{|P^-_T|}{1+|N^-_T|}\Bigg].
\$
Next, we define 
\$ 
M_\ell := \frac{|P_\ell^-|}{1+|N^-_\ell|}
\text{ and } 
\cG_\ell = \sigma\Big\{O_\ell, N^+_k, |N^-_\ell|, |P^+_\ell|, |P^-_\ell|,
\{(Z_i,A_i)\}_{i \in O_\ell \cup N^+_k}, \{X_i\}_{i\in U_\ell}\Big\}
\$
It can be checked $\{\cG_\ell\}_{\ell \ge k}$ is a filtration, and
that $M_\ell$ is $\cG_\ell$-measurable. It can also be observed that 
$\cF_\ell \subseteq \cG_\ell$, $\forall \ell \ge k$.
We shall then show that $\{M_\ell\}_{\ell \ge k}$ 
is a super-martingale sequence with respect to $\{\cG_\ell\}_{\ell\ge k}$, 
and then apply the optional stopping theorem to conclude the proof.

To see that $\{M_\ell\}_{\ell\ge k}$ is a super-martingale sequence, 
we fix $\ell \ge k$ and check the conditional expectation:
\@\label{eq:martingale}
\EE[M_{\ell +1} \given \cG_\ell] & = 
\EE\Bigg[\frac{|P^-_{\ell +1}|}{1+|N^-_{\ell +1}|} \bigggiven \cG_\ell\Bigg]\notag \\
& \stackrel{\textnormal{(a)}}{=} \EE\Bigg[\ind\big\{Y_{\pi(\ell+1)} \not\in \cC_{\pi(\ell+1)}\big\}\frac{|P^-_{\ell}|}{1+|N^-_{\ell}|} \bigggiven \cG_\ell\Bigg] 
+ \EE\Bigg[\ind\big\{Y_{\pi(\ell+1)} \in \cC_{\pi(\ell+1)}, A_{\pi(\ell+1)} = 0\big\}\frac{|P^-_{\ell}|}{|N^-_{\ell}|} \bigggiven \cG_\ell\Bigg] \notag \\
& \qquad \qquad + \EE\Bigg[\ind\big\{Y_{\pi(\ell+1)} \in \cC_{\pi(\ell+1)}, A_{\pi(\ell+1)} = 1\big\}\frac{|P^-_{\ell}|-1}{1+|N^-_{\ell}|} \bigggiven \cG_\ell\Bigg] \notag \\
& \stepb{=}\frac{|P^-_{\ell}|}{1+|N^-_{\ell}|} \cdot \PP\big(Y_{\pi(\ell+1)} \not\in \cC_{\pi(\ell+1)}\given \cG_\ell\big) 
+ \frac{|P^-_{\ell}|}{|N^-_{\ell}|} \cdot \PP\big(Y_{\pi(\ell+1)} \in \cC_{\pi(\ell+1)}, A_{\pi(\ell+1)} = 0 \given \cG_\ell\big)\notag \\
& \qquad \qquad + \frac{|P^-_{\ell}|-1}{1+|N^-_{\ell}|} \cdot
\PP\big(Y_{\pi(\ell+1)} \in \cC_{\pi(\ell+1)}, A_{\pi(\ell+1)} = 1\given \cG_\ell\big).
\@
Above, step (a) is by the definition of $P_\ell^-$ and $N^-_\ell$; 
step (b) is since $|P_\ell^-|$ and $|N^-_\ell|$ are $\cG_\ell$-measurable.

To proceed, we need the following lemma, which states that conditional on $\cG_\ell$, 
any of the unscreened {\em null} samples is equally likely to be selected as $\pi(\ell+1)$.
Its proof is deferred to Appendix~\ref{appd:proof_equal_prob}.
\begin{lemma}
\label{lemma:equal_prob} 
For any $\ell \ge k$, 
\$ 
& \PP(A_{\pi(\ell+1)}  = 0 \given \cG_\ell, Y_{\pi(\ell+1)} \in \cC_{\pi(\ell+1)}) 
= \frac{|N_\ell^-|}{|P_\ell^-| + |N_\ell^-|},\\
& \PP(A_{\pi(\ell+1)}  = 1 \given \cG_\ell, Y_{\pi(\ell+1)} \in \cC_{\pi(\ell+1)}) 
= \frac{|P_\ell^-|}{|P_\ell^-| + |N_\ell^-|}.
\$
\end{lemma}
Applying Lemma~\ref{lemma:equal_prob} to~\eqref{eq:martingale}, we have
\$
\eqref{eq:martingale} & = \frac{|P^-_{\ell}|}{1+|N^-_{\ell}|} \cdot \PP\big(Y_{\pi(\ell+1)} \not\in \cC_{\pi(\ell+1)}\given \cG_\ell\big)
+\frac{|P_\ell^-|}{|N_\ell^-|}\frac{|N^-_\ell|}{|N^-_\ell| + |P^-_\ell|} \cdot \PP(Y_{\pi(\ell+1)} \in \cC_{\pi(\ell+1)} \given \cG_{\ell}) \\
& \qquad \qquad + \frac{|P^-_\ell| -1 }{1+|N_\ell^-|} \frac{|P^-_\ell|}{|N^-_\ell| + |P^-_\ell|}\cdot \PP(Y_{\pi(\ell+1)} \in \cC_{\pi(\ell+1)} \given \cG_\ell)\\
& = \frac{|P^-_{\ell}|}{1+|N^-_{\ell}|} \cdot \PP\big(Y_{\pi(\ell+1)} \not\in \cC_{\pi(\ell+1)}\given \cG_\ell\big)
+ \bigg(\frac{|P_\ell^-|}{|N_\ell^-| + |P_\ell^-|} + \frac{|P_\ell^-|(|P_\ell^-| - 1)}{(1+|N_\ell^-|)(|N_\ell^-| + |P_\ell^-|)}\bigg)
\cdot \PP(Y_{\pi(\ell+1)} \in \cC_{\pi(\ell+1)} \given \cG_\ell)\\
&= \frac{|P^-_\ell|}{1+|N^-_\ell|}
 = M_\ell.
\$
We have now established that $\{M_\ell\}_{\ell \ge k}$ is a super-martingale sequence with respect to the filtration $\{\cG_\ell\}_{\ell\ge k}$.
Since $T$ is a stopping time with respect to the same filtration, we now 
apply the optional stopping theorem to conclude that 
$\EE[M_T] \le \EE[M_k]$. We finish the proof by 
bounding $\EE[M_k]$.


For notational convenience, define $n' := n-k$. Let $T_k^- := |P_k^-|+|N_k^-|$ denote the total number of null samples among the data not set aside in the initial training phase. We will condition on $T_k^-$ and show that \begin{equation}\label{bound-on-mk}
\EE[M_k \mid T_k^-] \leq \frac{m}{1+n'},
\end{equation} from which the result immediately follows. To show~\eqref{bound-on-mk}, observe that the exchangeability between the labeled and test data implies that $|P_k^-| \mid T_k^-\sim \text{Hypergeom}(m+n', T_k^-, m)$.\footnote{As a reminder, the first parameter is the total population size, the second is the number of units labeled ``success'' (in our case, the number of units which are null samples), and the third is the number of draws.} Consequently, inequality~\eqref{bound-on-mk} is a direct result of Lemma~\ref{hypergeom-lemma} below,
whose proof can be found in Appendix~\ref{appx:proof_hypergeom_lemma}.

\begin{lemma}\label{hypergeom-lemma}
    Let $X \sim \text{Hypergeom}(m+n', \kappa, m)$ for any non-negative integer $\kappa \leq m+n'$. Then \[\EE\left[\frac{X}{1+\kappa-X}\right] \leq \frac{m}{1+n'}\]
\end{lemma}

\subsection{Proof of Proposition~\ref{prop:diversity}}
\label{appd:proof_diversity_opt}
Throughout the proof, we use $\one$ to denote the vector of ones of length $\tilde n$
without explicit dependence on $\tilde n$.
Letting $\gamma_i = {\xi_i}/{\xi^\top \hat \delta}$, we rewrite the optimization problem 
in~\eqref{eq:diversity_opt} as 
\$
\min_{\gamma} ~&\gamma^\top \Theta \gamma,\\
\text{ s.t. } ~ &\gamma^\top \one = \frac{1}{1-\alpha},\\ 
& \gamma^\top \hat \delta = 1.
\$
The condition $\xi^\top \hat \delta > 0$ vanishes 
since $\gamma$ is agnostic to the sign of $\xi^\top \hat \delta$.
The Lagrangian of the above optimization problem is 
\$
\cL(\gamma, \lambda, \mu) = \gamma^\top \Theta \gamma + 
\lambda\Big(\gamma^\top \one - \frac{1}{1-\alpha} \Big) + \mu(\gamma^\top \hat \delta - 1),
\$
where $\lambda$ and $\mu$ are Lagrange multipliers. 
Letting $(\gamma^*,\lambda^*,\mu^*)$ be the optimal solution, 
the KKT condition yields that 
\$ 
\frac{\partial }{\partial \gamma}\cL(\gamma^*, \lambda^*, \mu^*) 
= 2\Theta \gamma^* + \lambda^* \one + \mu^* \hat \delta = 0
\Rightarrow \gamma^* = -\frac{1}{2}\Theta^{-1}(\lambda^* \one + \mu^* \hat \delta). 
\$
The other two KKT conditions are
\$
& (\gamma^*)^\top \hat \delta = -\frac{1}{2}(\lambda^* a + \mu^* b) = 1,\\
& (\gamma^*)^\top \one =  
-\frac{1}{2}(\lambda^* c+\mu^* a) =\frac{1}{1-\alpha} .
\$
The above yields 
\$
\mu^* = \frac{2}{bc - a^2} \Big(\frac{a}{1-\alpha} - c\Big), 
\text{ and }
\lambda^* = \frac{2}{bc - a^2} \Big(a - \frac{b}{1-\alpha}\Big).
\$ 
Putting things together, we have 
\$ 
\gamma^* = \frac{1}{bc-a^2} \bigg[\Big(\frac{b}{1-\alpha} - a\Big) \Theta^{-1}\one 
+ \Big(c - \frac{a}{1-\alpha}\Big)\Theta^{-1}\hat \delta \bigg].
\$
Recalling $\xi^* = \gamma^* \cdot (\xi^*)^\top \hat \delta$,  we have the desired result.
\section{Proof of Additional Lemmas}
\subsection{Proof of Lemma~\ref{lemma:equal_prob}}
\label{appd:proof_equal_prob}
To show the claim, we are to prove a stronger statement: for any $\ell \ge k$, and any 
$i \in P_\ell^- \cup N_\ell^-$, 
\@\label{eq:conditional_prob}
\PP\big(\pi(\ell+1) = i \given \cG_\ell, Y_{\pi(\ell+1)} \in \cC_{\pi(\ell+1)}\big) = 
\frac{1}{|P_\ell^-| + |N_\ell^-|}.
\@
Recall that by construction, $\pi(\ell+1)$ is a function of the information in $\cG_\ell$. Slightly abusing the notation, we write 
$\pi(\ell+1) := h_{\ell+1}(\cG_{\ell})$ for some 
measurable function $h_{\ell+1}$; we also define 
\$
\bar{\cG}_\ell = \big(\bar O_\ell, 
\bar N_k^+, |\bar N_\ell^-|, |\bar P_\ell^+|, |\bar P_\ell^-|,
\{(\bar Z_i, \bar A_i)\}_{i \in \bar O_\ell \cup \bar N^+_k }, 
\{\bar X_i\}_{i\in U_\ell}\big)
\$ 
and $\bar{P}_\ell^+$ denote 
the realization of $\cG_\ell$ and $P^+_\ell$, respectively.
Then 
\$ 
& \PP\big(\pi(\ell+1) = i \given \cG_\ell = \bar \cG_\ell, P^+_\ell = \bar P_\ell^+, Y_{\pi(\ell+1)} \in \cC_{\pi(\ell+1)}\big)\\ 
= ~&\frac{\PP\big(h_{\ell+1}(\bar \cG_\ell)= i, \cG_\ell = \bar \cG_\ell, 
P^+_\ell = \bar P^+_\ell, Y_{\pi(\ell+1)} \in \cC_{\pi(\ell+1)}\big)}
{\PP\big(\cG_\ell = \bar \cG_\ell, P^+_\ell = \bar P_\ell, Y_{\pi(\ell+1)} \in \cC_{\pi(\ell+1)}\big)}.
\$
For any $j \in \bar P_\ell^- \cup \bar N^-_\ell$, note that both $\cG_\ell$ 
and $P^+_\ell$ are invariant to the swapping of $Z_i$ and $Z_j$.
Also by exchangeability, swapping $Z_i$ and 
$Z_j$ does not change the joint distribution of the data points, and therefore
\$ 
& \PP\big(h_{\ell+1}(\bar \cG_\ell)= i, \cG_\ell = \bar \cG_\ell, 
P^+_\ell = \bar P^+_\ell, Y_i \in \cC_i\big) \\
= ~& \PP\big(h_{\ell+1}(\bar \cG_\ell)= j, \cG_\ell = \bar \cG_\ell, 
P^+_\ell = \bar P^+_\ell, Y_j \in \cC_j\big)\\
=~& \frac{1}{|\bar P^-_\ell|+|\bar N^-_\ell|} 
\sum_{j \in \bar P_\ell^- \cup \bar N_\ell^-}
\PP\big(\pi(\ell+1)= j, \cG_\ell = \bar \cG_\ell, 
P^+_\ell = \bar P^+_\ell, Y_{\pi(\ell+1)} \in \cC_{\pi(\ell+1)}\big)\\
=~& \frac{1}{|\bar P^-_\ell|+|\bar N^-_\ell|} 
\PP\big(\cG_\ell = \bar \cG_\ell,
P^+_\ell = \bar P^+_\ell, Y_{\pi(\ell+1)} \in \cC_{\pi(\ell+1)}\big),
\$ 
which implies~\eqref{eq:conditional_prob}.
Using~\eqref{eq:conditional_prob}, we finally arrive at
\$ 
& \PP\big(A_{\pi(\ell+1)} = 1 \given \cG_\ell, Y_{\pi(\ell+1)} \in \cC_{\pi(\ell+1)}\big)  \\
= ~ &\EE\Big[\PP\big(A_{\pi(\ell+1)} = 1 \given \cG_\ell, Y_{\pi(\ell+1)} \in \cC_{\pi(\ell+1)}, P_\ell^+\big) 
\biggiven \cG_\ell, Y_{\pi(\ell+1)} \in \cC_{\pi(\ell+1)}\Big]\\
& = \frac{|N^-_\ell|}{|N^-_\ell|+|P^-_\ell|},
\$
completing the proof.

\subsection{Proof of Lemma~\ref{hypergeom-lemma}}
\label{appx:proof_hypergeom_lemma}

    We have that 
    \begin{align*}
        \EE\left[\frac{X}{1+\kappa-X}\right] &= \sum_{x=\max(1,\kappa-n')}^{\min(m,\kappa)} \frac{x}{1+\kappa-x} \frac{\binom{\kappa}{x}\binom{m+n'-\kappa}{m-x}}{\binom{m+n'}{m}}\\
        &= \sum_{x=\max(1,\kappa-n')}^{\min(m,\kappa)} \frac{\binom{\kappa}{x-1}\binom{m+n'-\kappa}{m-x}}{\binom{m+n'}{m}}\\
        &= \sum_{x=\max(1,\kappa-n')}^{\min(m,\kappa)} \frac{\binom{\kappa}{x-1}\binom{m+n'-\kappa}{(m-1)-(x-1)}}{\binom{m+n'}{m-1} \cdot \frac{n'+1}{m}}\\
        &= \frac{m}{n'+1} \sum_{x=\max(0,\kappa-(n'+1))}^{\min(m,\kappa)-1} \frac{\binom{\kappa}{x}\binom{m+n'-\kappa}{m-1-x}}{\binom{m+n'}{m-1}}\\
        &\leq \frac{m}{n'+1} \sum_{x=\max(0,\kappa-(n'+1))}^{\min(m-1,\kappa)} \frac{\binom{\kappa}{x}\binom{m+n'-\kappa}{m-1-x}}{\binom{m+n'}{m-1}}\\
        &= \frac{m}{n'+1},
    \end{align*} where the final equality is due to the fact that the summation in the penultimate line is $1$ as it is of the PMF of the $\text{Hypergeom}(m+n', \kappa, m-1)$ distribution across its support.

\section{Additional simulation results}
\label{appd:simulation}
This section contains additional simulation results. In addition to the 
data generating process described in Section~\ref{sec:simulation} (referred to 
as Setting 1 hereafter), we investigate the performance of our proposed method 
under several other settings adapted from~\citet{jin2023selection}. 
Supposing $Y_i \sim \cN(\mu(X_i), \sigma(X_i)^2)$, we consider 
the following choices of $\mu(\cdot)$ and $\sigma(\cdot)$: 
\begin{itemize}
\item Setting 2: $\mu(x) = 5(x_1 x_2 + e^{x_4-1})$, $\sigma(x) = 1.5\cdot \sigma$;
\item Setting 3: $\mu(x) = 5(x_1 x_2 + e^{x_4-1})$, $\sigma(x) = \sigma \cdot (5.5-|\mu(x)|)/2$;
\item Setting 4: $\mu(x) = 5(x_1 x_2 + e^{x_4-1})$, $\sigma(x) = \sigma\cdot 0.25\mu(x)^2 \ind\{|\mu(x)|<2\} 
+ \sigma\cdot 0.5|\mu(x)|\ind\{|\mu(x)|\ge 1\}$; 
\item Setting 5: $\mu(x) = 2(x_1x_2 + x_3^2 +e^{x_4-1}-1)$, $\sigma(x) = 1.5\cdot \sigma$. 
\end{itemize}

Section~\ref{appd:sim_indiv_base_model} contains the additional results of 
ACS with different individual base models under Settings 2-5.
Section~\ref{appd:sim_model_selection} contains the results of 
ACS with model selection under Settings 2-5. Section~\ref{appd:sim_diversity}
provides the results of diversity-aware ACS with different choices of 
$\lambda$, while Section~\ref{appd:sim_k} investigates the effect of 
the choice of $k$ on the performance of ACS.

\subsection{Additional results with individual base models}
\begin{figure}[h!]
\centering
\includegraphics[width=\textwidth]{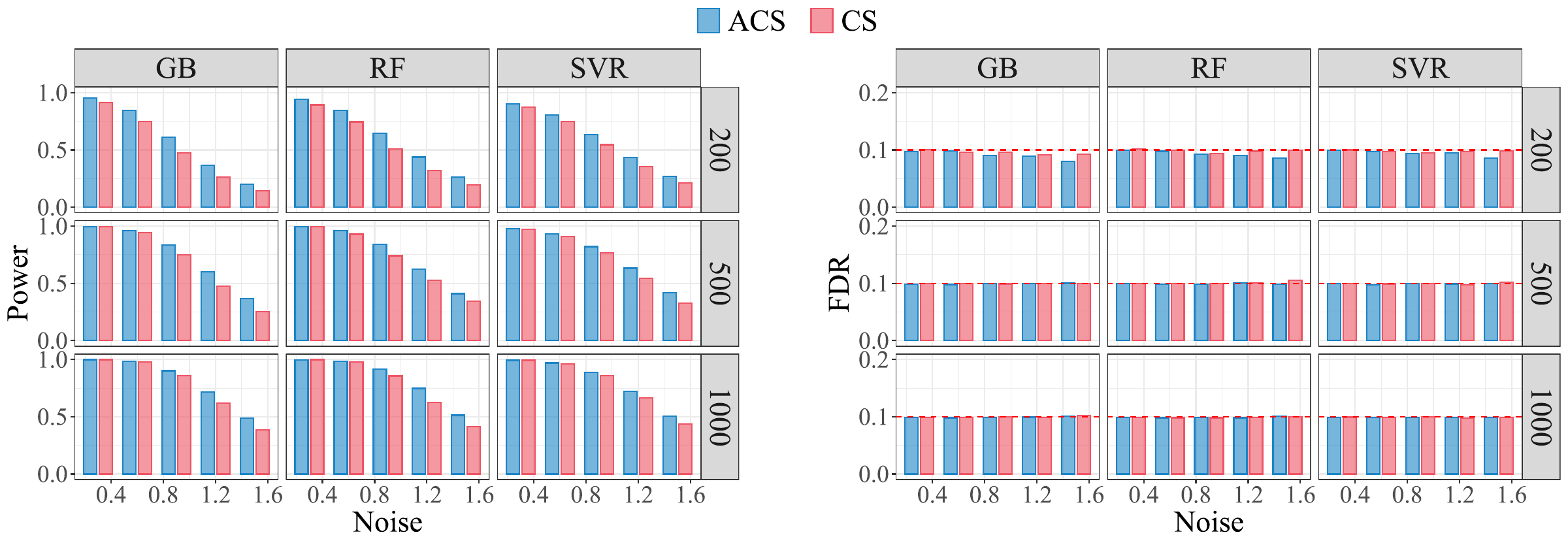} 
\caption{Realized power (left) and FDR (right) of ACS and CS 
under simulation Setting 2. The other details are the same as in Figure~\ref{fig:set1_simulation}.}
\label{fig:set2_simulation}
\end{figure}

\begin{figure}[h!]
\centering
\includegraphics[width=\textwidth]{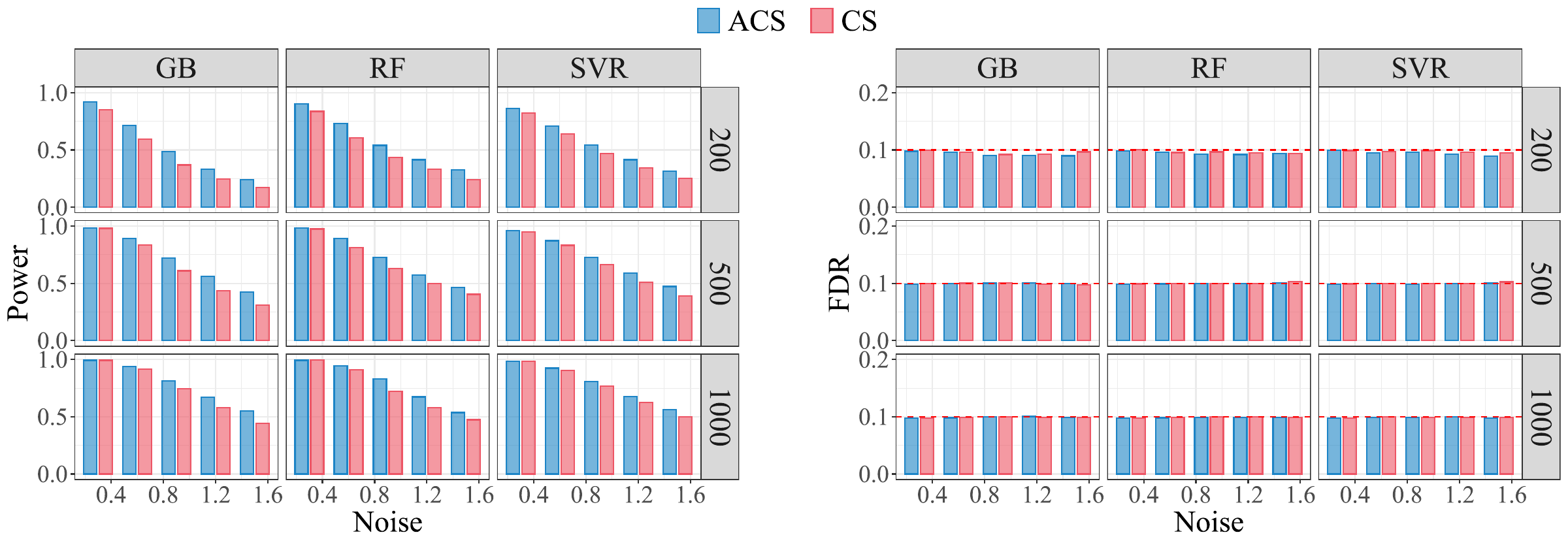} 
\caption{Realized power (left) and FDR (right) of ACS and CS 
under simulation Setting 3. The other details are the same as in Figure~\ref{fig:set1_simulation}.}
\label{fig:set3_simulation}
\end{figure}

\begin{figure}[h!]
\centering
\includegraphics[width=\textwidth]{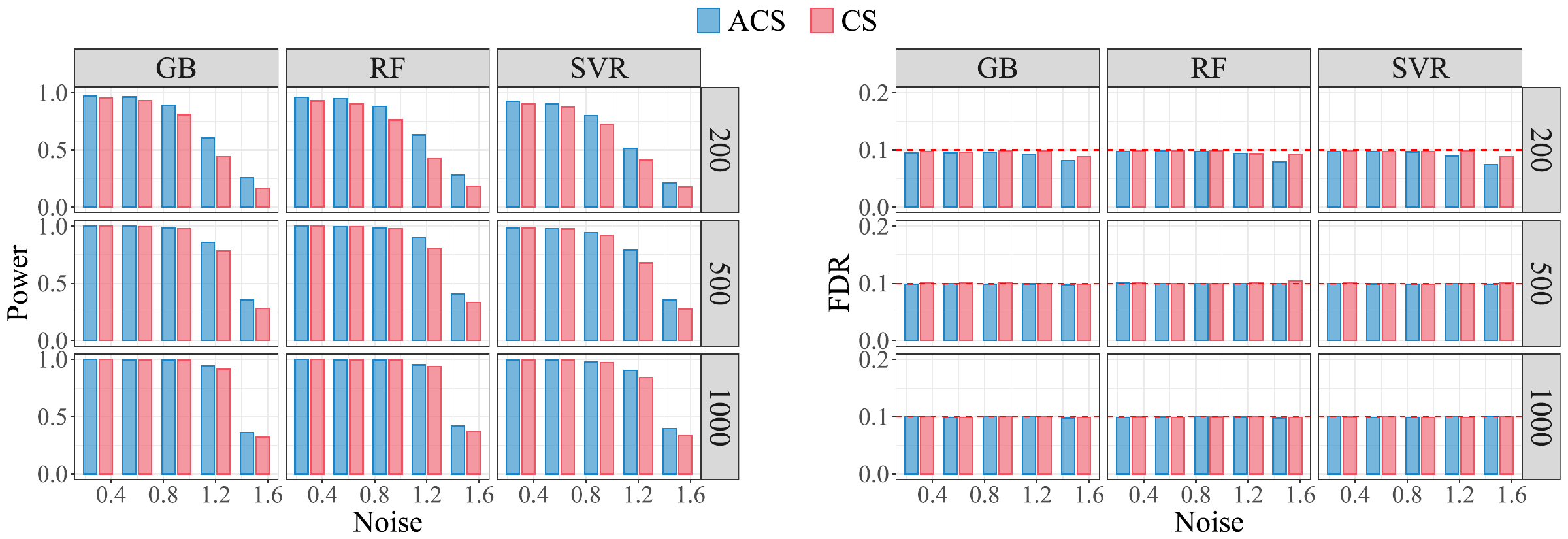} 
\caption{Realized power (left) and FDR (right) of ACS and CS 
under simulation Setting 4. The other details are the same as in Figure~\ref{fig:set1_simulation}.}
\label{fig:set4_simulation}
\end{figure}

\begin{figure}[h!]
\centering
\includegraphics[width=\textwidth]{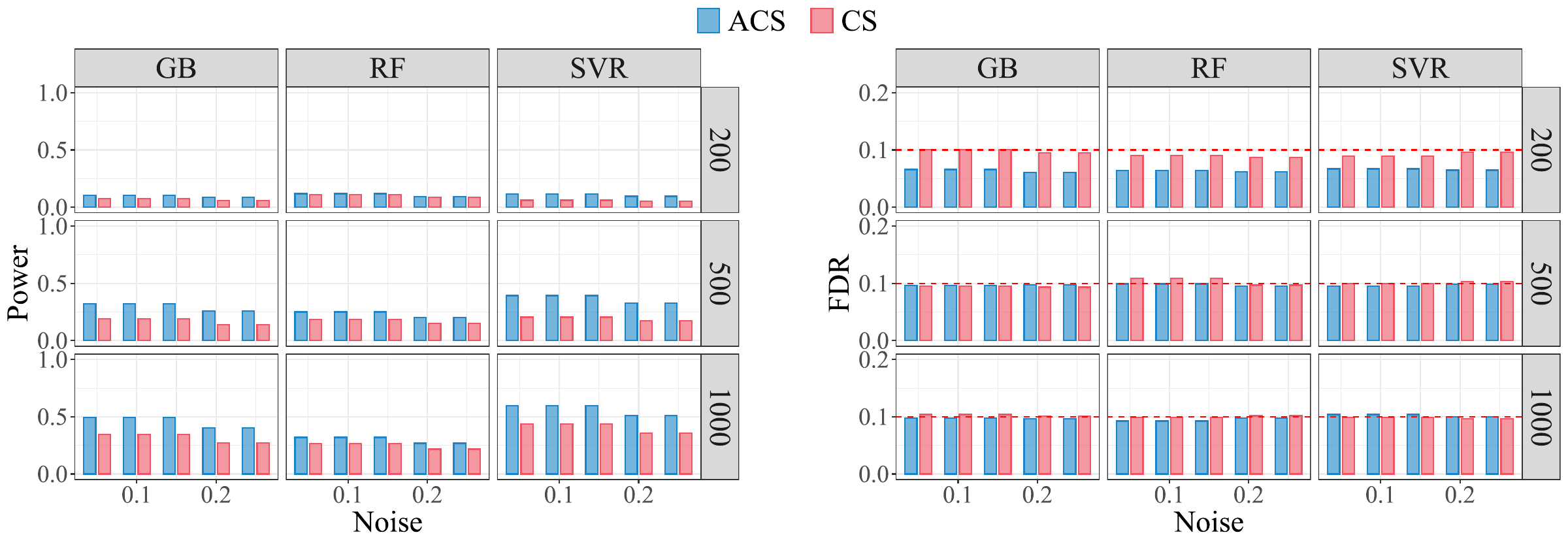} 
\caption{Realized power (left) and FDR (right) of ACS and CS 
under simulation Setting 5. The other details are the same as in Figure~\ref{fig:set1_simulation}.}
\label{fig:set5_simulation}
\end{figure}

Figures~\ref{fig:set2_simulation}--\ref{fig:set5_simulation} 
show the realized power and FDR of ACS and CS under Settings 2-5.
The same story holds as in Setting 1: ACS uniformly outperforms CS in terms of
realized power for diferrent individual base models or different samples sizes, 
while maintaining the FDR at the nominal level.

\label{appd:sim_indiv_base_model}

\subsection{Model selection}
\begin{figure}[h!]
\centering
\includegraphics[width=0.8\textwidth]{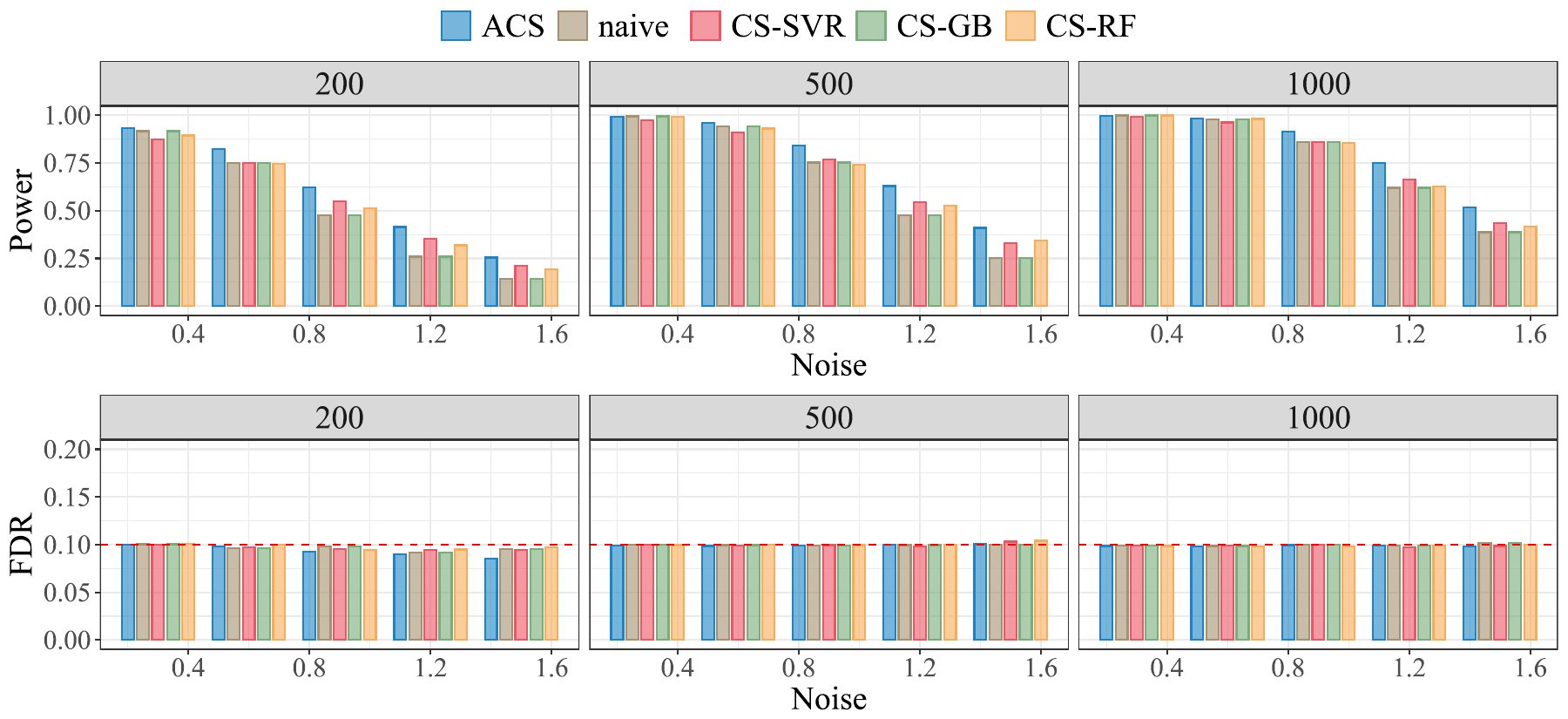} 
\caption{Realized power (left) and FDR (right) of ACS with adaptive model selection, 
the naive method, CS-SCR, CS-RF, and CS-GB as a function of the noise level $\sigma$
under simulation Setting 2. The other details are the same as in Figure~\ref{fig:ensemble_set1_simulation}.}
\label{fig:ensemble_set2_simulation}
\end{figure}

\begin{figure}[h!]
\centering
\includegraphics[width=0.8\textwidth]{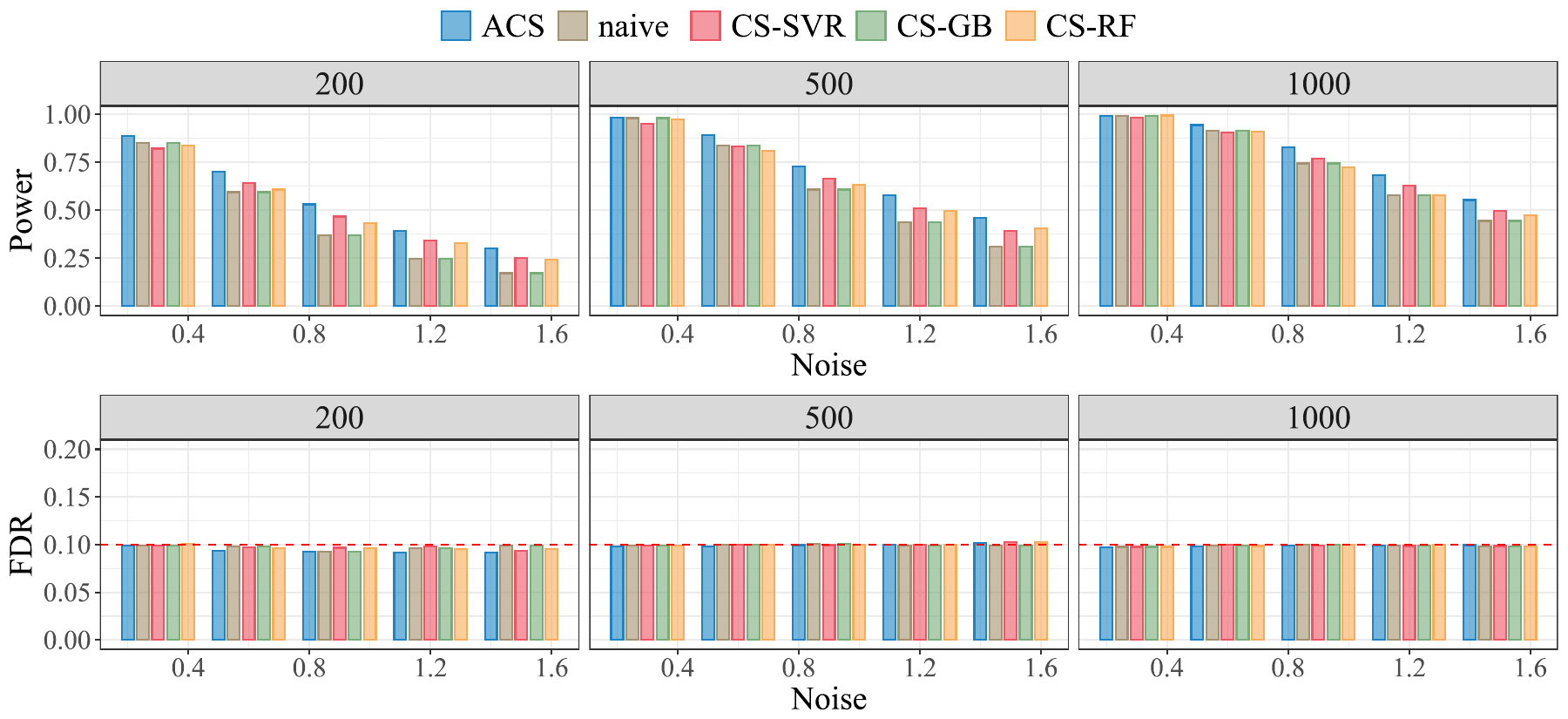} 
\caption{Realized power (left) and FDR (right) of ACS with adaptive model selection, 
the naive method, CS-SCR, CS-RF, and CS-GB as a function of the noise level $\sigma$
under simulation Setting 3. The other details are the same as in Figure~\ref{fig:ensemble_set1_simulation}.}
\label{fig:ensemble_set3_simulation}
\end{figure}

\begin{figure}[h!]
\centering
\includegraphics[width=0.8\textwidth]{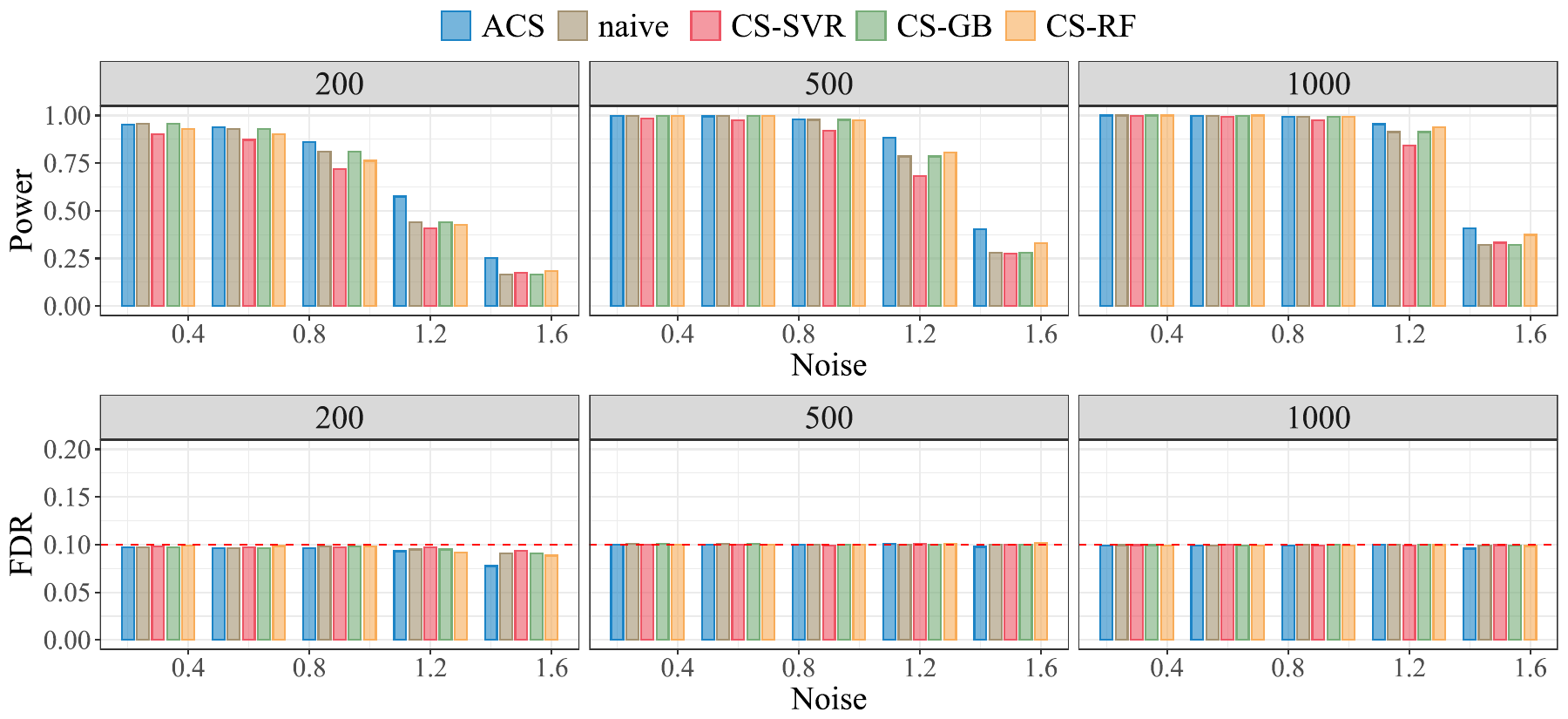} 
\caption{Realized power (left) and FDR (right) of ACS with adaptive model selection, 
the naive method, CS-SCR, CS-RF, and CS-GB as a function of the noise level $\sigma$
under simulation Setting 4. The other details are the same as in Figure~\ref{fig:ensemble_set1_simulation}.}
\label{fig:ensemble_set4_simulation}
\end{figure}

\begin{figure}[h!]
\centering
\includegraphics[width=0.8\textwidth]{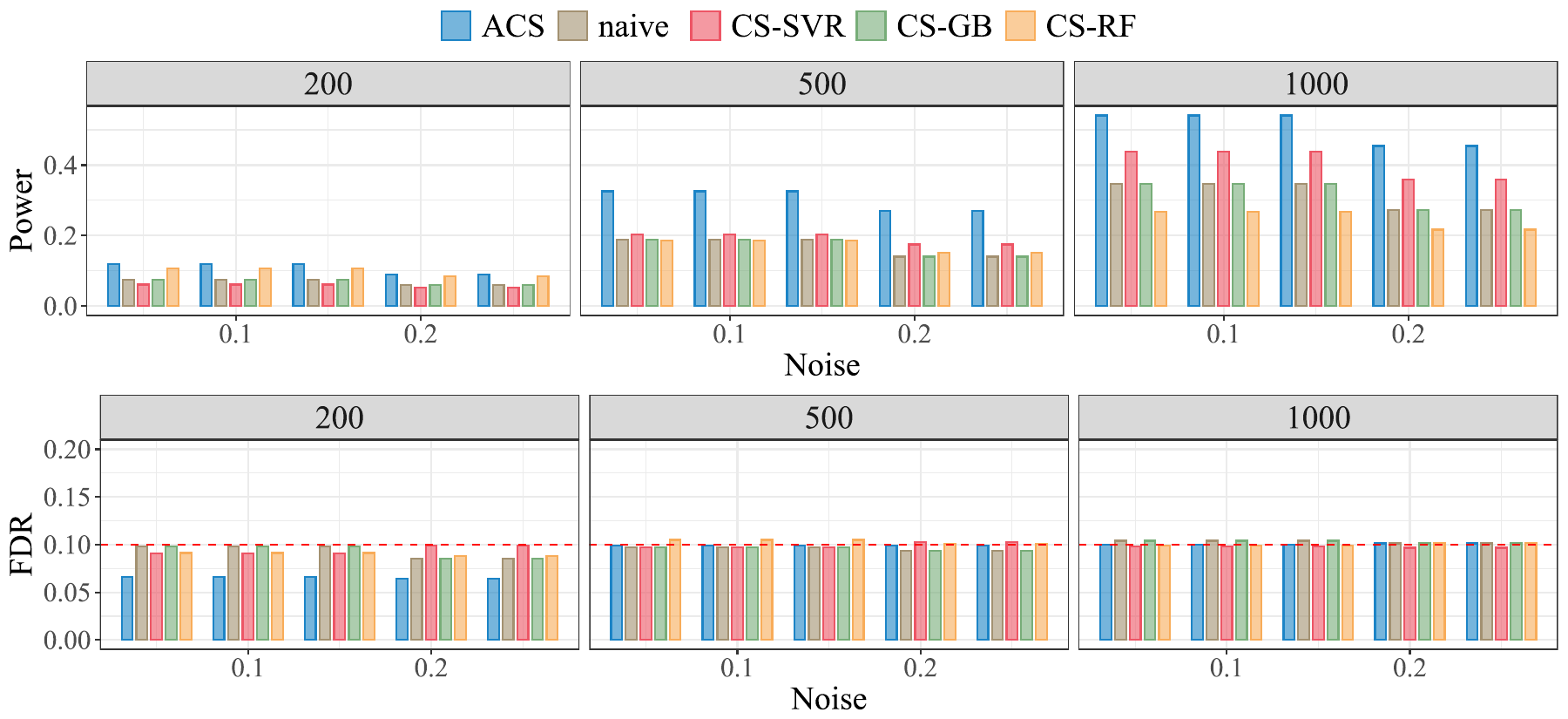} 
\caption{Realized power (left) and FDR (right) of ACS with adaptive model selection, 
the naive method, CS-SCR, CS-RF, and CS-GB as a function of the noise level $\sigma$
under simulation Setting 5. The other details are the same as in Figure~\ref{fig:ensemble_set1_simulation}.}
\label{fig:ensemble_set5_simulation}
\end{figure}

Figures~\ref{fig:ensemble_set2_simulation}--\ref{fig:ensemble_set5_simulation}
plot the realized power and FDR of ACS with adaptive model selection, the naive method,
CS-SCR, CS-RF, and CS-GB under Settings 2-5. In all settings considered, ACS dominates 
the best base models, as well as the naive method, showing its ability to automatically 
pick the best model. Once again, the FDR of ACS is well controlled at the nominal level.

\label{appd:sim_model_selection}

\subsection{Diversity-aware selection}
\label{appd:sim_diversity}
\begin{figure}[h!]
    \centering
    \includegraphics[width=.6\textwidth]{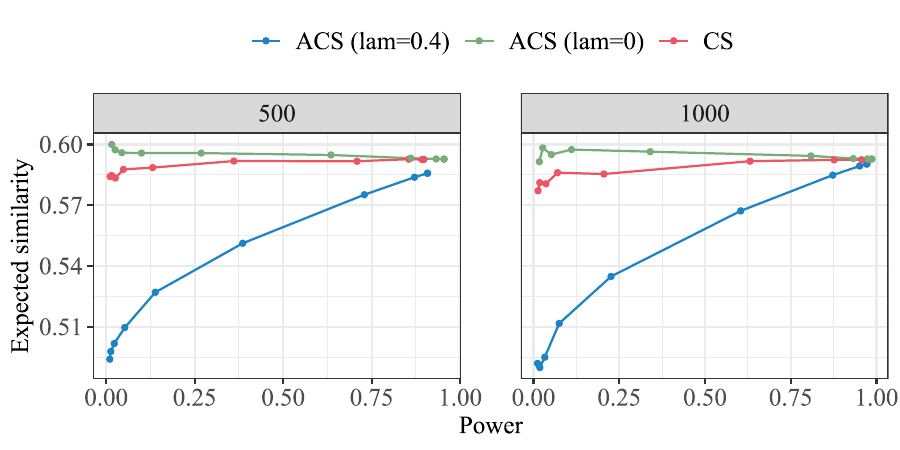}
    \caption{$\widehat{\text{ES}}$ as a function of $\widehat{\text{Power}}$ for 
diversity-aware ACS with $\lambda = 0.4$, ACS, and CS under Setting 1. 
The other details are the same as in Figure~\ref{fig:div_set_es_1_lam_0.3_curve}.}
    \label{fig:div_set_es_1_lam_0.4_curve}
\end{figure}

\begin{figure}[h!]
    \centering
    \includegraphics[width=0.8\textwidth]{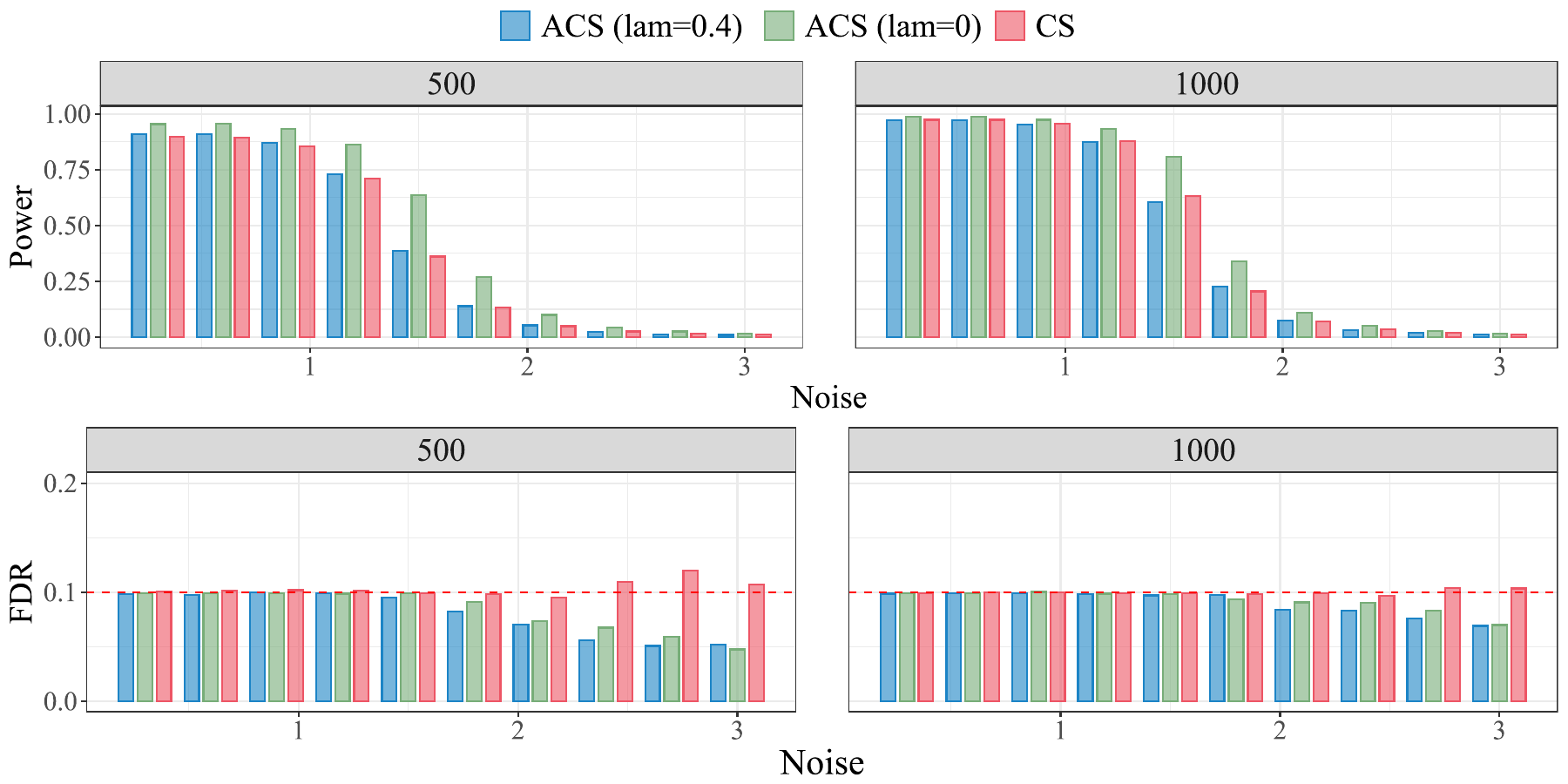}
    \caption{Realized power (left) and FDR (right) of diversity-aware ACS with $\lambda = 0.4$,
ACS, and CS as a function of the noise level $\sigma$ under simulation Setting 1. The other details are the same as in Figure~\ref{fig:div_set_es_1_lam_0.3}.}
    \label{fig:div_set_es_1_lam_0.4}
\end{figure}

\begin{figure}[h!]
    \centering
    \includegraphics[width=.6\textwidth]{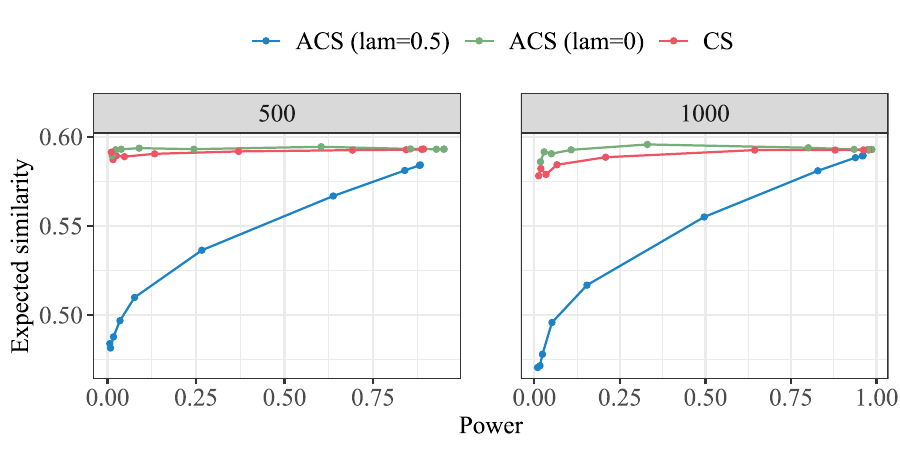}
    \caption{$\widehat{\text{ES}}$ as a function of $\widehat{\text{Power}}$ for 
diversity-aware ACS with $\lambda = 0.5$, ACS, and CS. The other details are the same as in Figure~\ref{fig:div_set_es_1_lam_0.3_curve}.}
    \label{fig:div_set_es_1_lam_0.5_curve}
\end{figure}

\begin{figure}[h!]
    \centering
    \includegraphics[width=0.8\textwidth]{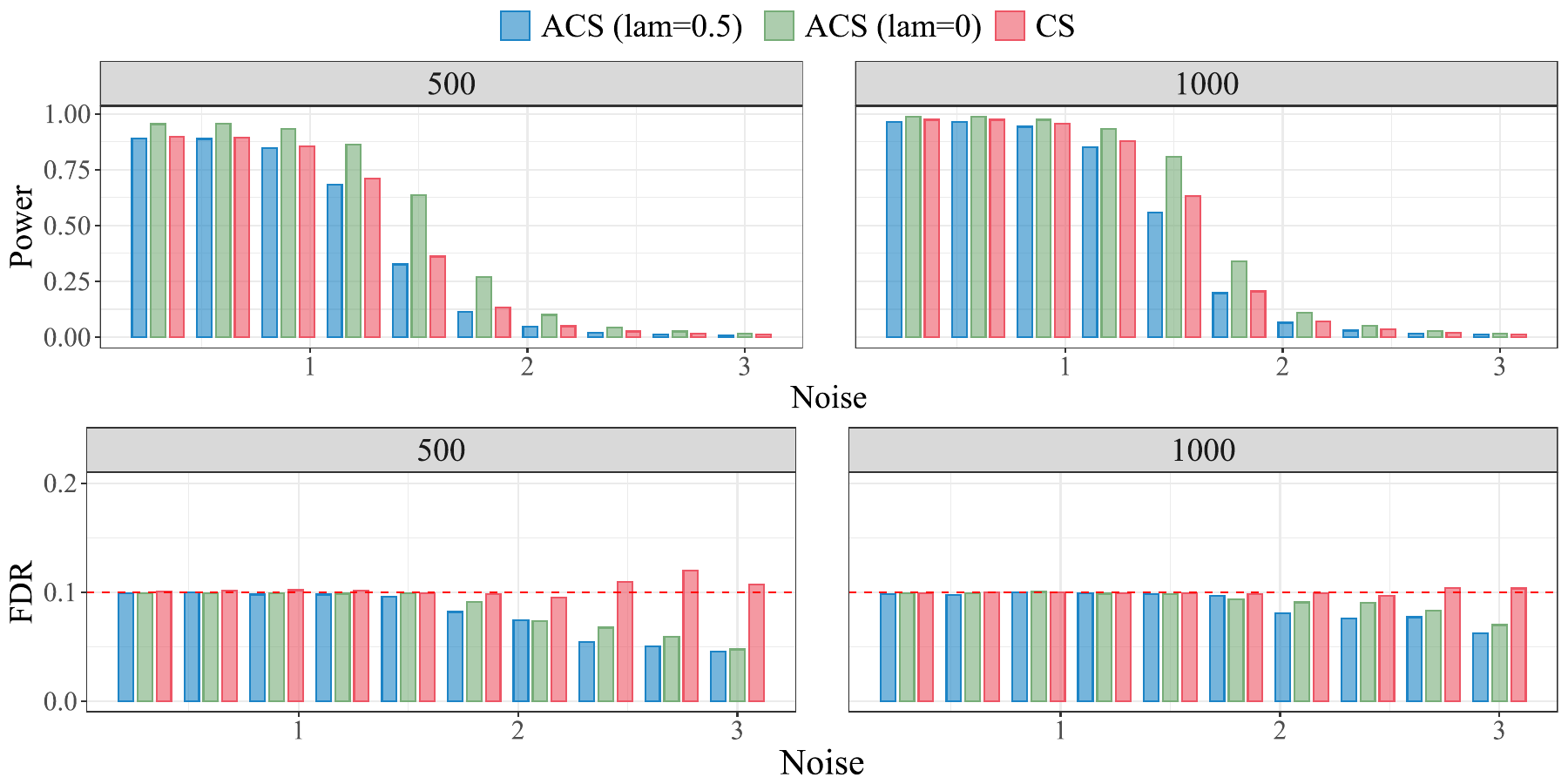}
    \caption{Realized power (left) and FDR (right) of diversity-aware ACS with $\lambda = 0.5$,
ACS, and CS as a function of the noise level $\sigma$ under simulation Setting 1. The other details are the same as in Figure~\ref{fig:div_set_es_1_lam_0.3}.}
    \label{fig:div_set_es_1_lam_0.5}
\end{figure}

This section contains the results of diversity-aware ACS under Setting 1 
with other choices of $\lambda$. Figures~\ref{fig:div_set_es_1_lam_0.4_curve} 
and~\ref{fig:div_set_es_1_lam_0.5_curve}
plot the $\widehat{\text{ES}}$ as a function of $\widehat{\text{Power}}$; compared with 
the results in Figure~\ref{fig:div_set_es_1_lam_0.3_curve}, we can see that increasing 
$\lambda$ leads to a decrease in the $\widehat{\text{ES}}$, reflecting a higher weight 
on the diversity term. Figures~\ref{fig:div_set_es_1_lam_0.4} and~\ref{fig:div_set_es_1_lam_0.5}
show the realized power and FDR of diversity-aware ACS with $\lambda = 0.4$ and $\lambda = 0.5$,
respectively. The power of ACS slightly decreases as $\lambda$ increases---this is plausibly due to that 
the procedure refrains from making selections that are similar to the ones already made. 
The FDR remains controlled across different values of $\lambda$.

\subsection{The effect of $k$}
\label{appd:sim_k}
\begin{figure}[h!]
    \centering
    \includegraphics[width = 0.8\textwidth]{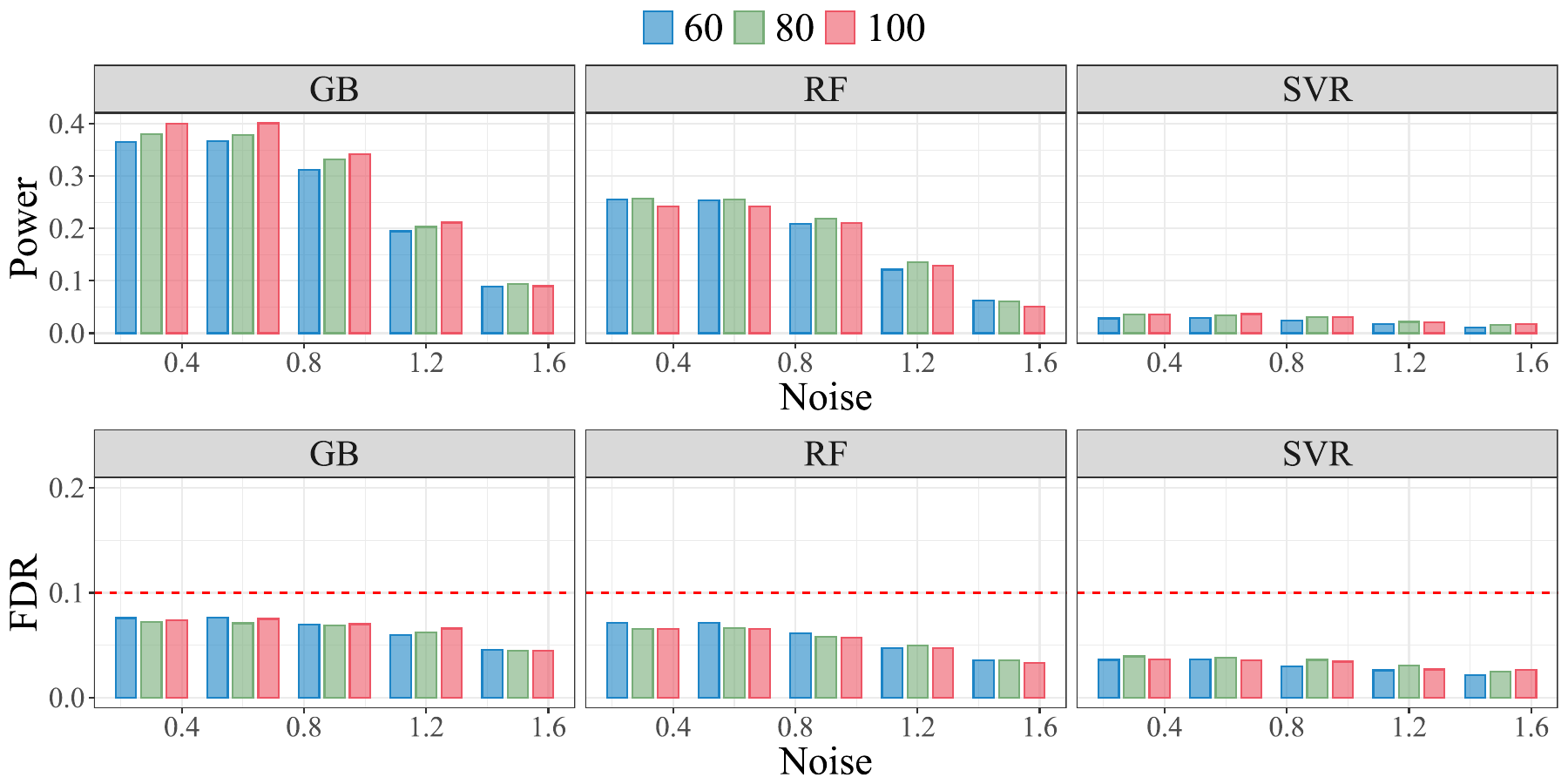}
    \caption{Realized power (left) and FDR (right) of ACS with $200$
    labeled samples and $k\in \{60,80,100\}$ under simulation Setting 1. 
    The other details are the same as in Figure~\ref{fig:set1_simulation}.}
    \label{fig:set1_simulation_k}
\end{figure}

Under Setting 1, we implement ACS with $200$ labeled samples and  
$k$ (the number of initial samples used to train the model)
ranging in $\{60,80,100\}$. 
The results are visualized in Figure~\ref{fig:set1_simulation_k}. The results do 
not differ significantly across different values of $k$.

\section{Additional real data results}

\subsection{Additional details and results for LLM deployment}\label{sec:additional_llm}

In this section, we present more implementation details for Section~\ref{sec:real_data} and additional results with a smaller model \textbf{OPT-13B}.

\subsubsection{Feature engineering for prediction model training}
Following~\citet{gui2024conformal}, to build the prediction model $\hat g$, we use a separate tuning set $\cD_{\rm tune}$ for feature engineering.
More specifically, we focus on quantities that are extracted from foundation model outputs.
As the quality of the outputs can be affected by the difficulty of the task suggested by the prompt as well as LLM's capability in this specific task--which can be revealed in both the likelihood of the output and the confidence of the model itself, we consider the following quantities as the available covariates $\tilde X_j$:
\begin{itemize}
	\item Self-evaluation likelihoods~\citep{kadavath2022language,lin2023generating}. The self-evaluation score (\texttt{Self\_Eval}), or \texttt{P(True)}, is the model's own evaluation for the generated answer, which is obtained using the same prompt in \cite{kadavath2022language} as follows:
	\begin{tcolorbox}
	\texttt{[story]\\
	Question: [question]\\
	Here are some brainstormed ideas: [few\_shots examples]\\
	Possible Answer: [generated answer]
	Is the possible answer:\\
	(A) True\\
	(B) False\\
	The possible answer is: (
	}    
	\end{tcolorbox}
	The self-evaluation score is then defined as the output probability associated with the generated answer ``\texttt{A)}''.
	\item Input uncertainty scores~\citep{kuhn2023semantic, lin2023generating}. To quantify the uncertainty—or difficulty—of an input, \cite{kuhn2023semantic} proposed generating $M$ outputs for each prompt. If the responses are consistent, the input is considered less uncertain; conversely, high variability in the generated answers indicates greater uncertainty. Specifically, the selected features include lexical similarity (\texttt{Lexical\_Sim}), measured by the \texttt{rouge-L} score between generated answers. To capture semantic variation, we apply a natural language inference (NLI) classifier to group the $M$ answers into semantically consistent clusters, from which we derive two features: the number of semantic groups (\texttt{Num\_Sets}) and the semantic entropy (\texttt{SE}). Consistent with \cite{kuhn2023semantic, lin2023generating}, we employ a pre-trained DeBERTa-large model \citep{he2020deberta} as the NLI predictor.
	\item Output confidence scores~\citep{lin2023generating,gui2024conformal}. High likelihoods assigned by large language models (LLMs) do not necessarily correspond to high-quality outputs, as these likelihoods can be poorly calibrated \citep{zhao2021calibrate,xiong2023can}. To address this issue, \cite{lin2023generating} introduced confidence measures derived from the structural similarity among $M$ generated outputs. These measures include the eigenvalues of the graph Laplacian (\texttt{EigV}), the average pairwise distances computed via the degree matrix (\texttt{Deg}), and the eccentricity (\texttt{Ecc}), which incorporates embedding information from each generation. Each metric is computed with respect to a defined similarity function. Following the notation in \cite{lin2023generating}, we append suffixes \texttt{J}, \texttt{E}, and \texttt{C} to indicate the use of Jaccard similarity, NLI-based entailment scores, and NLI-based contradiction scores, respectively.
\end{itemize}
With the aforementioned scores, the covariates $\tilde X_j$ for question answering tasks is a $13$-dimensional vector consisting of \texttt{Self\_Eval}, \texttt{Lexical\_Sim}, \texttt{Num\_Sets}, \texttt{SE}, \texttt{EigV(C/E/J)}, \texttt{Ecc(C/E/J)}, and \texttt{Deg(C/E/J)}. In the X-ray report generation task, as the self-evaluation pipeline is not applicable for image inputs, we define $\tilde X_j$ as a $12$-dimensional vector with remaining scores.

\subsubsection{Additional experimental results}
As the results with \textbf{LLaMA-2-13B-chat} on \textbf{CoQA} are presented in Section~\ref{sec:real_data}, as a comparison with a different base LLM and a different dataset, we present results with a smaller model \textbf{OPT-13B} \citep{zhang2023opt} and results on the \textbf{TriviaQA} dataset in this section. The data splitting, ACS, and out-of-sample evaluation pipeline remains the same with Section~\ref{sec:real_data}.

\begin{figure}[h]
\centering
\includegraphics[width=\textwidth]{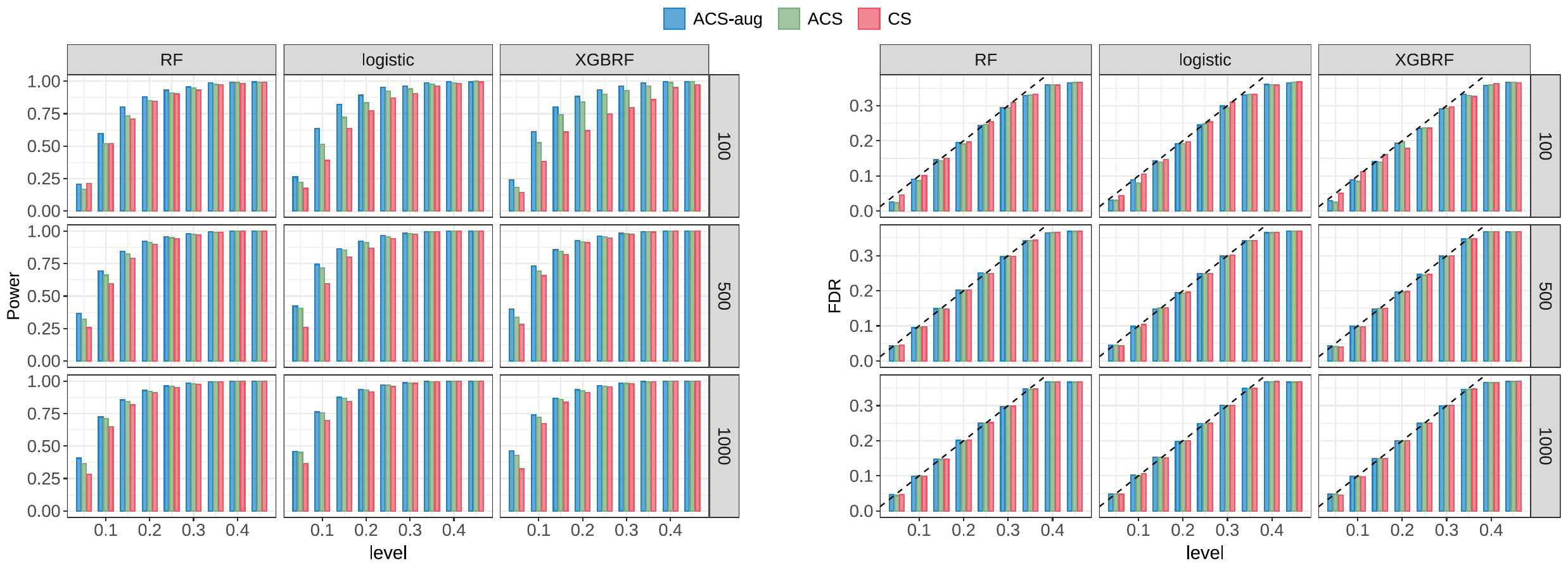} 
\caption{Realized power (left) and FDR (right) of ACS with new labels (ACS-aug), ACS without new labels
(ACS), and CS: LLaMA-2-13B-chat + TriviaQA dataset. Details are otherwise the same as Figure~\ref{fig:aug-lm-coqa}.}
\label{fig:aug-lm-triviaqa}
\end{figure}

\begin{figure}[h]
\centering
\includegraphics[width=\textwidth]{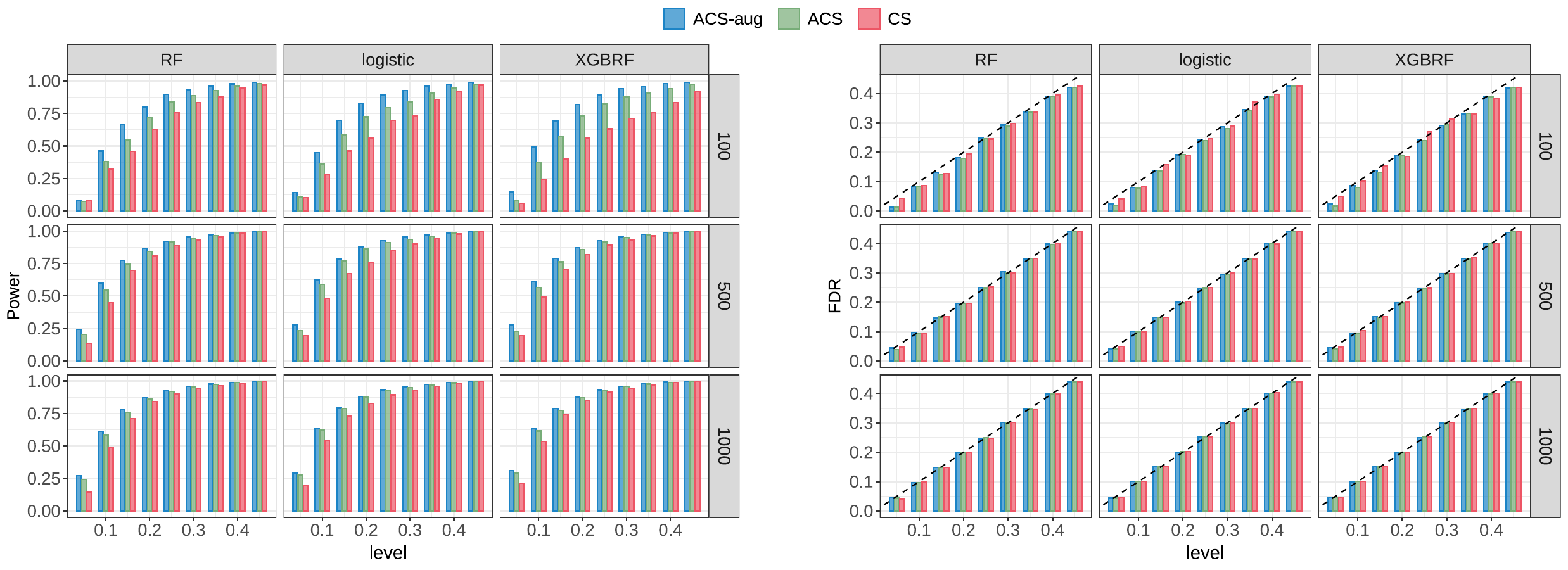} 
\caption{Realized power (left) and FDR (right) of ACS with new labels (ACS-aug), ACS without new labels
(ACS), and CS: OPT-13B + CoQA dataset. 
The target FDR level is $\alpha = 0.1$ 
and the results are averaged over $200$ independent simulations.}
\label{fig:aug-opt-coqa}
\end{figure}

\begin{figure}[h]
\centering
\includegraphics[width=\textwidth]{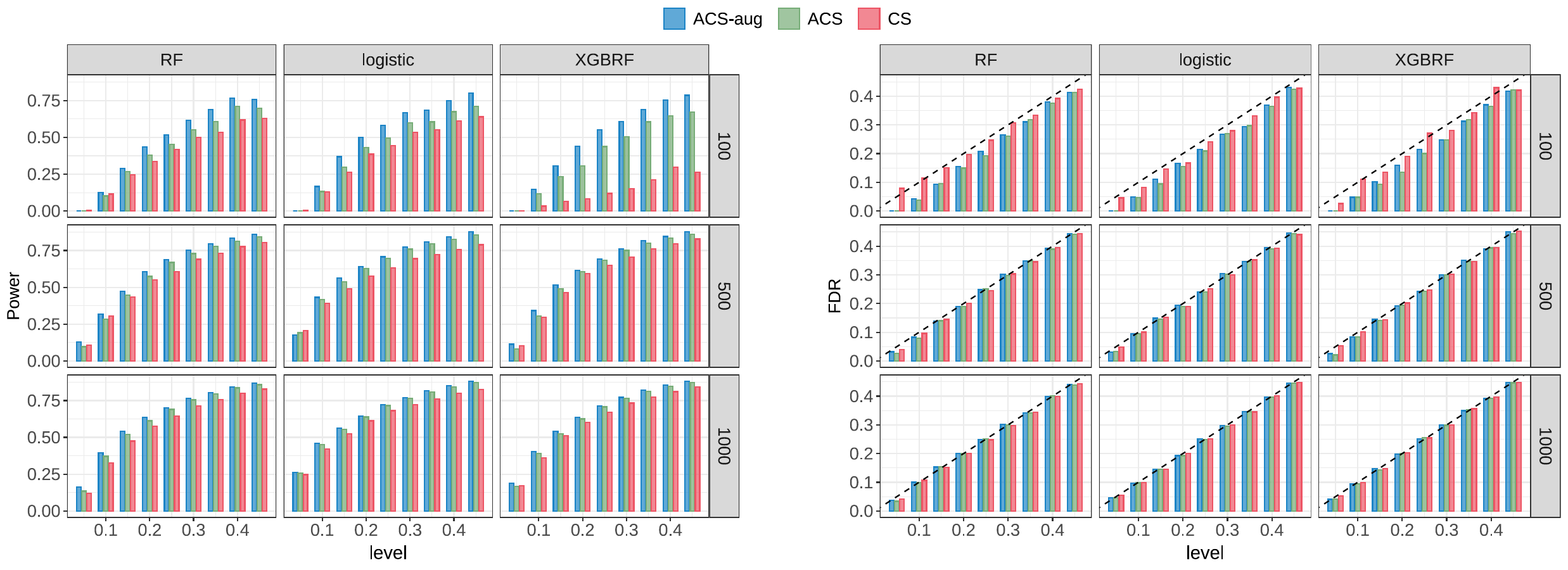} 
\caption{Realized power (left) and FDR (right) of ACS with new labels (ACS-aug), ACS without new labels
(ACS), and CS: OPT-13B + TriviaQA dataset. 
The target FDR level is $\alpha = 0.1$ 
and the results are averaged over $200$ independent simulations.}
\label{fig:aug-opt-triviaqa}
\end{figure}

\paragraph{CS, ACS, and ACS with new labels incorporation.}
By fixing $\cG \in \{\texttt{logistic}, \texttt{RF}, \texttt{XGBRF}\}$, Experimental results for ACS with new labels (ACS-aug), ACS without label incorporation (ACS), 
and CS on CoQA and TriviaQA datasets with OPT-13B as the base LLM are presented in Figure~\ref{fig:aug-opt-coqa} and Figure~\ref{fig:aug-opt-triviaqa}.
Figure~\ref{fig:aug-lm-triviaqa} presents the additional results of LLaMA-2-13B-chat on TriviaQA.
FDR is controlled for each method for different nominal FDR levels $\alpha$, but when the nominal level is small, the power of ACS and ACS-aug is much higher than that of CS. Similar to the results with LLaMA-2-13B-chat, the difference in the power of ACS, ACS-aug and CS is even more outstanding with a small sample size and a data-consuming predictive model (e.g., \texttt{XGBRF}).

\begin{figure}[h]
\centering
\includegraphics[width=\textwidth]{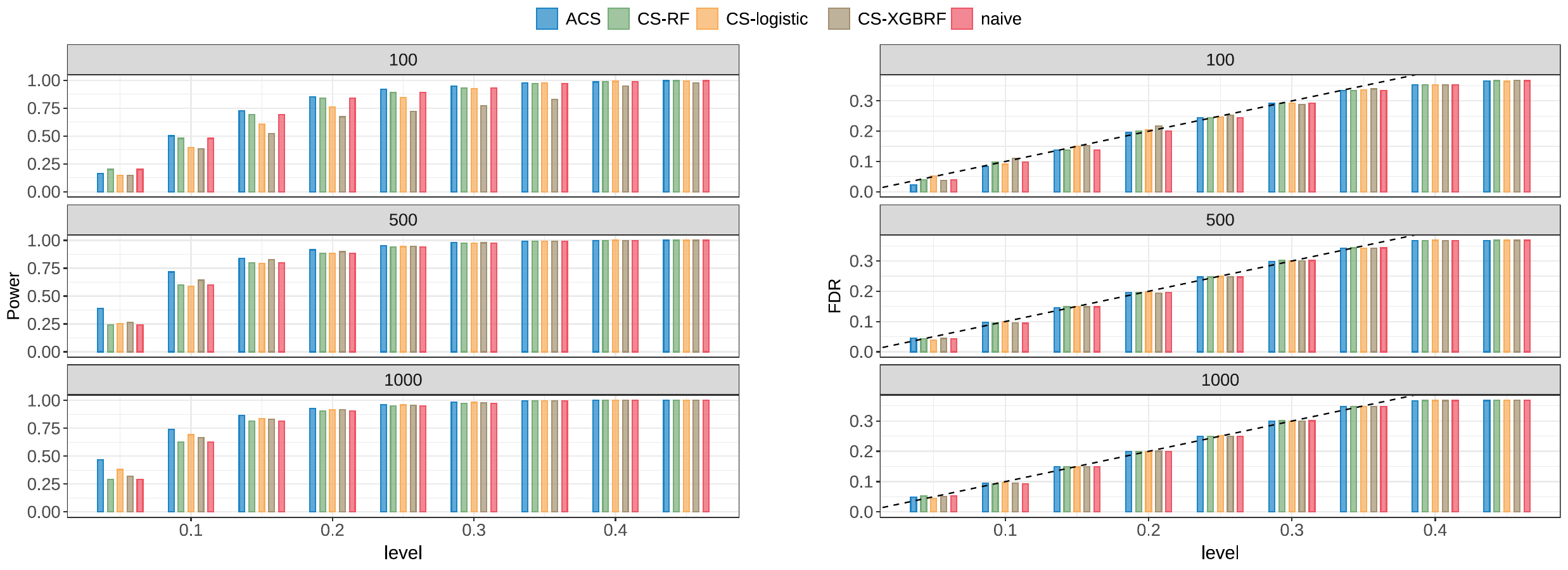} 
\caption{Realized power (left) and FDR (right) of ACS with adaptive model selection, 
the naive method, CS-SVR, CS-RF, and CS-GB: LLaMA-2-13B-chat + TriviaQA dataset. Each subplot corresponds to the size $N$ of labeled samples.
The target FDR level is $\alpha = 0.1$ 
and the results are averaged over $200$ independent simulations.}
\label{fig:en-lm-triviaqa}
\end{figure}

\begin{figure}[h]
\centering
\includegraphics[width=\textwidth]{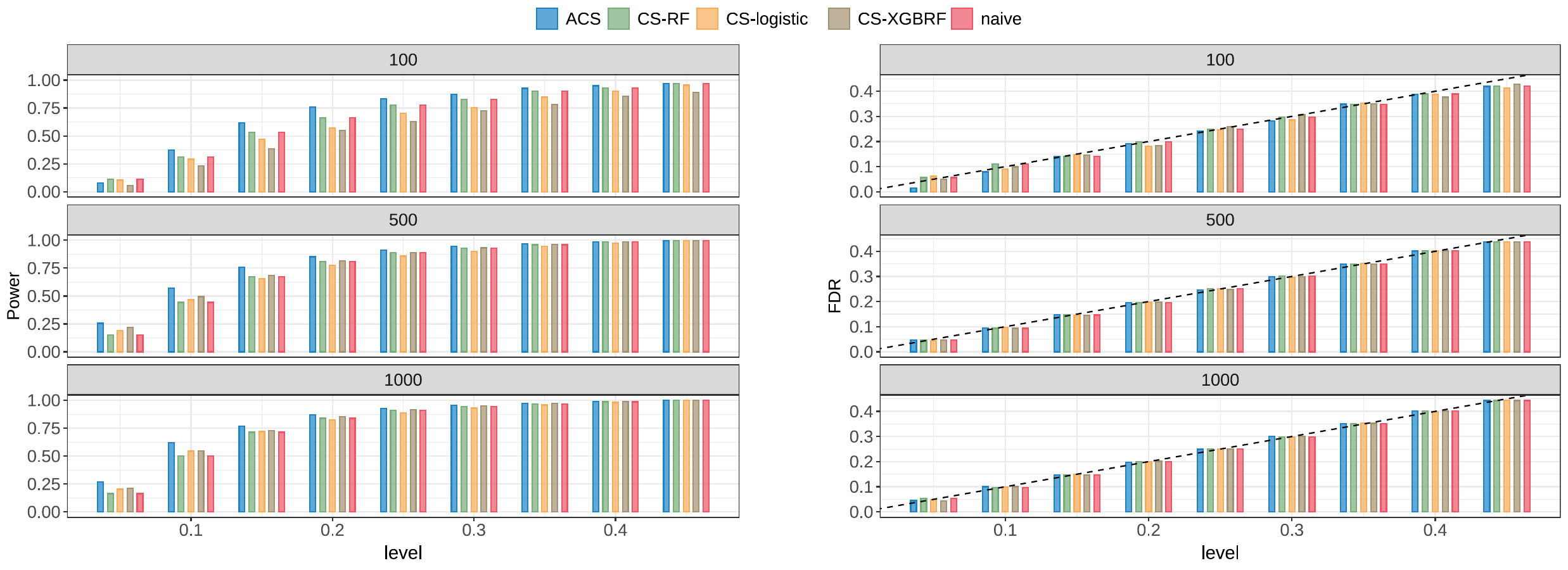} 
\caption{Realized power (left) and FDR (right) of ACS with adaptive model selection, 
the naive method, CS-SVR, CS-RF, and CS-GB: OPT-13B + CoQA dataset.
The target FDR level is $\alpha = 0.1$ 
and the results are averaged over $200$ independent simulations.}
\label{fig:en-opt-coqa}
\end{figure}

\begin{figure}[h]
\centering
\includegraphics[width=\textwidth]{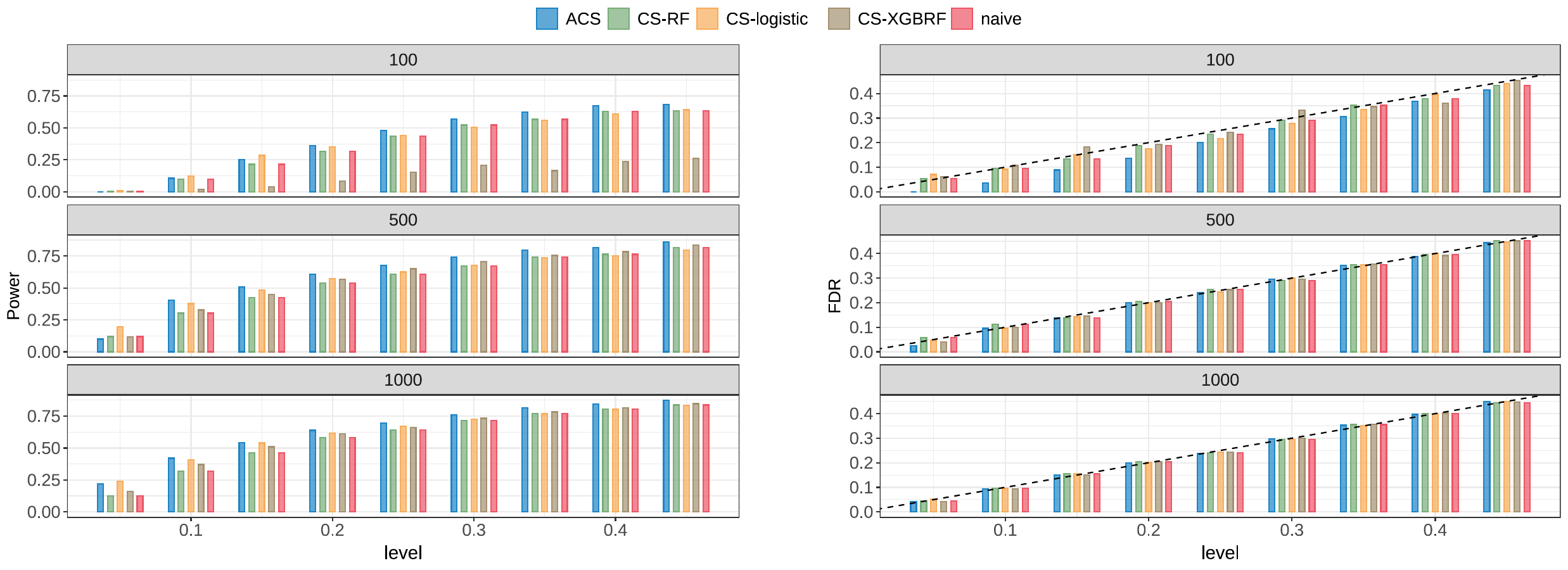} 
\caption{Realized power (left) and FDR (right) of ACS with adaptive model selection, 
the naive method, CS-SVR, CS-RF, and CS-GB: OPT-13B + TriviaQA dataset.
The target FDR level is $\alpha = 0.1$ 
and the results are averaged over $200$ independent simulations.}
\label{fig:en-opt-triviaqa}
\end{figure}

\paragraph{ACS with adaptive model selection.}
Using the same pipeline described in Section~\ref{sec:adaptive_model_selection}, we compare ACS with adaptive model selection against several baselines: the naive method, CS-logistic, CS-RF, and CS-XGBRF. Figure~\ref{fig:en-lm-triviaqa} presents the realized power and FDR with LLaMA-2-13B-chat on TriviaQA, and Figures~\ref{fig:en-opt-coqa} and~\ref{fig:en-opt-triviaqa} report the realized power and FDR of OPT-13B for CoQA and TriviaQA, respectively. Across both datasets, ACS with adaptive model selection consistently achieves the highest power while maintaining FDR control at the target level. This advantage becomes especially evident when the sample size is small or the target FDR level is particularly stringent.

\subsection{Additional drug discovery results}\label{appendix:drug-discovery}
\begin{figure}[h]
\centering
\includegraphics[width=\textwidth]{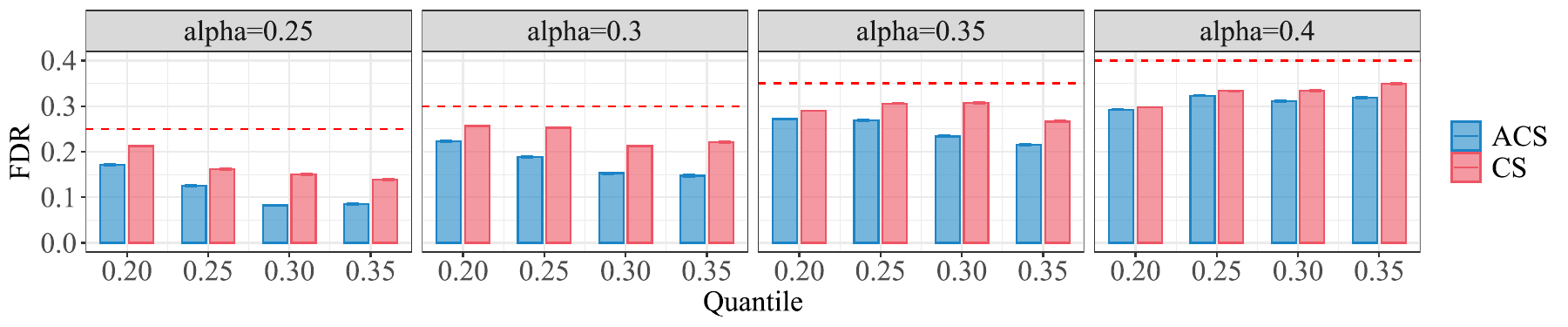} 
\caption{FDR of ACS (blue) compared to CS (red) for various quantile values $q$ and nominal FDR levels $\alpha$ on real drug discovery dataset. Horizontal lines denote the nominal level $\alpha$.}
\label{fig:drug-fdr}
\end{figure}

Figure~\ref{fig:drug-fdr} shows the FDR of ACS and CS in the same setting as described in Section~\ref{sec:drug_discovery}. Both methods control FDR conservatively.

\end{document}